\documentclass[10pt,preprint2]{aastex}

\usepackage{wasysym}
\usepackage{graphicx}
\usepackage{subeqn}

\begin{document}
\title{Outcomes and Duration of Tidal Evolution in a Star-Planet-Moon System}
\author{Takashi Sasaki} \affil{Department of Physics} \affil{University of Idaho}
\affil{Moscow, ID 83844-0903} \affil{ResearcherID:B-4470-2011  } \email{tsasaki@vandals.uidaho.edu}
\author{Jason W. Barnes} \affil{Department of Physics} \affil{University of Idaho}
\affil{Moscow, ID 83844-0903} \affil{ResearcherID:  B-1284-2009} \email{jwbarnes@uidaho.edu}
\author{David P. O'Brien} \affil{Planetary Science Institute} \affil{1700 East Fort Lowell, Suite 106}
\affil{Tucson, AZ 85719-2395} \email{obrien@psi.edu}
\newpage

\begin{abstract}
We formulated tidal decay lifetimes for hypothetical moons orbiting extrasolar planets with both lunar and stellar tides. Previous work neglected the effect of lunar tides on planet rotation, and are therefore applicable only to systems in which the moon's mass is much less than that of the planet. This work, in contrast, can be applied to the relatively large moons that might be detected around newly-discovered Neptune-mass and super-Earth planets. We conclude that  moons are more stable when the planet/moon systems are further from the parent star, the planets are heavier, or the parent stars are lighter. Inclusion of lunar tides allows for significantly longer lifetimes for a massive moon relative to prior formulations. We expect that the semi-major axis of the planet hosting the first detected exomoon around a G-type star is 0.4-0.6 AU and is 0.2-0.4 AU for an M-type star.
\end{abstract}

\keywords{celestial mechanics--planetary systems-planet and satellites:general}

\section{INTRODUCTION}
The first discovery of an extrasolar planet in orbit around a main-sequence star was made by \citet{1995Natur.378..355M}.
Since then, more than 700 extrasolar planets\footnote{http://exoplanet.eu/} have been discovered.
Although extrasolar moons have not yet been detected, they almost certainly exist.
Most of the planets in our solar system have satellites.
Even Pluto, though no longer officially a planet, has three moons \citep{2006Natur.439..943W}.
It is likely that the mechanisms for moon formation in our solar system (impact, capture, and coaccretion) prevail beyond it\citep{2011arXiv1108.4070M}.

The Earth's obliquity, or axial tilt, is stabilized by the Moon \citep{1993Natur.361..615L}.
Mars, on the other hand, has relatively small satellites, and its obliquity changes chaotically, fluctuating on a 100,000-year timescale \citep{1993Natur.361..608L}.
Stable obliquity in its star's habitable zone may be necessary for a planet to support life.
An Earth-size planet with no moon, or a relatively small one, may be subject to large fluctuations in obliquity.
In such a case, favorable conditions may not last long enough for life to become established.
In the same way, orbital longevity is required for any life form to have time to become established.
Hence, the prospects for habitable planets may hinge on moons \citep{2000rewc.book.....W}; but see also \citep{2012Icar..217...77L}.

In 2005, \citet{2005ApJ...634..625R} discovered Gliese 876 d, the first super-Earth around a main sequence star.
To date more than thirty super-Earths have been discovered\footnote{http://exoplanet.eu/}.
The discovery of Kepler 22-b in the habitable zone gives rise to the possibility of life beyond our Solar System \citep{2012ApJ...745..120B}.
It is important to know the lifetime of moons orbiting super-Earths in the habitable zone: while the planet might be unsuited to the evolution of life, its moons might be.
Moons with masses of at least one third $M_{\bigoplus}$, and orbiting around gas giant planets in the habitable zone may have habitable environments \citep{1997Natur.385..234W}.
The moon's orbital stability plays a role in habitability as well.
Clearly, if the moon leaves orbit, it will probably leave the habitable zone.

Although extrasolar moons have not yet been found, several methods to detect them have been investigated.
After \citet{2010ApJ...712L.125K}, the following methods can detect extrasolar moons:
\begin{enumerate}
\item Transit timing variations  \citep{1999A&AS..134..553S, 2005MNRAS.359..567A, 2005Sci...307.1288H}.
\item Transit duration variations \citep{2009MNRAS.396.1797K}.
\item Light curve distortions \citep{2006A&A...450..395S}.
\item Planet-moon eclipses \citep{2007A&A...464.1133C}.
\item Microlensing \citep{2008ApJ...684..684H}.
\item Pulsar timing \citep{2008ApJ...685L.153L}.
\item Distortion of the Rossiter-McLaughlin effect of a transiting planet \citep{2009EM&P..105..385S}.
\end{enumerate}
Considering the speed at which observational instrumentation has developed, it is only a matter of time before extrasolar moons are discovered.


Tidal torque is important to the long-term orbital stability of extrasolar moons.
A binary system can be in tidal equilibrium only if coplanarity (the equatorial planes of the planet and moon coincide with the orbital plane), circularity (of the orbit), and corotation (the rotation periods of the planet and moon are equal to the revolution period) have been fulfilled. Further, stability occurs only if the orbital angular momentum exceeds the sum of the spin angular momenta of the planet and moon by more than a factor of three \citep{1980A&A....92..167H}.

\citet{1973ApJ...180..307C} studied the stability of these equilibria only with respect to coplanarity and circularity.
He pointed out that in a planet-moon system with lunar\footnote{In this paper, we use ``lunar" as the adjective of any moons, not just the Moon} tides, there are three possible evolutionary states.

Counselman state (i):the semi-major axis of the moon's orbit tidally evolves inward until the moon hits the planet. Example:Phobos around Mars.

Counselman state (ii):the semi-major axis of the moon's orbit tidally evolves outward until the moon escapes from the planet.
No solar system examples are available. But this result would happen to the Earth-Moon if Earth's present rotation rate were doubled.

Counselman state (iii):lunar orbital and planetary spin angular velocities enter mutual resonance and are kept commensurate by tidal forces. Example:Pluto and Charon. This Counselman state is static, while state (i) and (ii) are evolutionary.

Here, we consider a star-planet-moon system with stellar tides.
Although they did not consider the effects of lunar tides or maximum distance from the planet,
\cite{1973MNRAS.164...21W} examined the impact of solar tides on planetary rotation in a limited star-planet-moon system.

\citet{2002ApJ...575.1087B} considered a similar case, considering the maximum distance of the moon but neglecting the lunar tide's effect on planetary rotation.
They found just two possible final states:the moon may either hit the planet or escape from it.

In this paper, we consider a star-planet-moon system with both stellar and lunar tides, and lunar maximum distance from the planet.
Stellar and lunar tides both affect planetary spin, whereas stellar and lunar tides affect planet and moon orbits, respectively.
We do not consider the effect of stellar tides on the moon's rotation.
Stellar tides should sap angular momentum from the system but this effect is less important if the mass of the planet is at least ten times greater than the mass of the moon.
We apply tidal theory and set up a system of differential equations that govern the planetary rotational rate and orbital mean motion as well as the orbital mean motion of the moon.
The system of differential equations is solved  numerically.
Finally, a formula for the length of time the moon will stably orbit is found.
We then apply this result to hypothetical extrasolar planet moon systems.

\section{THEORY} \label{section:THEORY}
In this paper, we use standard tidal evolution theory with the constant Q approach \citep{1966Icar....5..375G}, along with the following assumptions:
\begin{enumerate}
\item The spin angular momentum of the planet is parallel to the orbital angular momenta of both the moon about the planet and the planet about the star;
        i.e. the planet has $0^\circ$ obliquity, the moon orbits in the planet's equatorial plane, and the planet and moon motions are prograde.
\item The total angular momentum, that is the sum of the moon's orbital angular momentum and the planet's rotational and orbital angular momenta, is constant.
        We neglect the orbital angular momentum of the moon about the star and the moon's rotational angular momentum.
\item The moon's orbit about the planet and the planet's orbit about the star are circular.
\item The moon is less (at most $\sim$ 1/10) massive than the planet and the planet is also less (at most $\sim$ 1/10) massive than the star.
\item The star's spin angular momentum is not considered nor are the planet's tides on the star or the star's tides on the moon.
\item The specific dissipation function of the planet, $Q_{p}$, is independent of the tidal forcing frequency and does not change as a function of time.
\end{enumerate}

Planetary $Q_{p}$ falls into two groups.
The first group has values of $Q_{p}$ that range from 10 to 500.
The terrestrial planets and satellites of the Jovian planets are in this group.
The other group has $Q_{p}$ values greater than $10^4$.
The Jovian planets are in this group.
In the case of the Earth, tidal dissipation is due to friction between the tidally generated currents and the ocean floor and occurs mostly in shallow seas.
For Mercury and Venus, tidal dissipation is driven by viscous dissipation within the bulk planetary interior.
The mechanism for tidal dissipation within giant planets remains unknown.

In that case, the torque on the planet due to the moon $\tau_{p-m}$ is given by \citet{2002ApJ...575.1087B};\citet{1966Icar....5..375G}; \citet{SolarSystemDynamics} in Chapter 4

\begin{equation}\label{eq:eq1}
\tau_{p-m}=-\frac{3}{2}\frac{k_{2p}GM^{2}_{m}R^{5}_{p}}{Q_{p}a^{6}_{m}}{\rm sgn}(\Omega_{p}-n_{m}),
\end{equation}
where $\Omega_{p}$ is the rotational rate of the planet,
$k_{2p}$ is the tidal Love number of the planet,
$G$ is the gravitational constant,
$R_{p}$ is the radius of the planet,
$M_{m}$ is the mass of the moon,
$a_{m}$ is the semimajor axis of the moon's orbit, and
$n_{m}$ is the orbital mean motion of the moon.
The function sgn$(\Omega_{p}-n_{m})$ is 1 if $(\Omega_{p}-n_{m})$ is positive, -1 if $(\Omega_{p}-n_{m})$ is negative, and undefined if $(\Omega_{p}-n_{m})=0$.

Similarly, the torque on the planet due to the star $\tau_{p-s}$ is
\begin{equation}\label{eq:eq2}
\tau_{p-s}=-\frac{3}{2}\frac{k_{2p}GM^{2}_{s}R^{5}_{p}}{Q_{p}a^{6}_{p}}{\rm sgn}(\Omega_{p}-n_{p}),
\end{equation}
where $M_{s}$ is the mass of the star, $a_{p}$ is the semimajor axis of the planet's orbit, $n_{p}$ is the orbital mean motion of the planet.


For the spin angular momentum of the planet,
\begin{equation}\label{eq:eq3}
I_{p}\frac{d\Omega_{p}}{dt}=\frac{dL_{pspin}}{dt}=\tau_{p-m}+\tau_{p-s},
\end{equation}
where the planet's rotational moment of inertia $I_{p}=\alpha M_{p}R^2_{p}$. $\alpha$ is the moment of inertia constant.
For Earth and Jupiter, $\alpha$'s are 0.3308 and 0.254, respectively \citep{dePater.Lissauer}.

The change in orbital angular momenta of the moon about the planet and the planet about the star, by Newton's Third Law, are equal and opposite the moon's and star's torques on the planet, respectively:
\begin{equation}\label{eq:eq4}
\frac{dL_{m}}{dt}=\tau_{m-p}=-\tau_{p-m}
\end{equation}
and
\begin{equation}\label{eq:eq5}
\frac{dL_{p}}{dt}=\tau_{s-p}=-\tau_{p-s},
\end{equation}
where $L_{m}=M_{m}a^{2}_{m}n_{m}$ and $L_{s}=M_{p}a^{2}_{p}n_{p}$.

Using Kepler's Third Law, $n^{2}_{m}a^{3}_{m}\approx GM_{p}$ and $n^{2}_{p}a^{3}_{p}\approx GM_{s}$ because we assume that $M_{m}\ll M_{p}$ and $M_{p}\ll M_{s}$. This allows us to derive these expressions for $L_{m}$ and $L_{p}$
\begin{equation}
L_{m}=\frac{M_{m}(GM_{p})^{2/3}}{n_{m}^{1/3}}
\end{equation}
and
\begin{equation}
L_{p}=\frac{M_{p}(GM_{s})^{2/3}}{n_{p}^{1/3}}.
\end{equation}


To determine how $n_{m}$ and $n_{p}$ vary with time,
we take the derivative of $L_{m}$ and $L_{p}$ with respect to $t$, set the results equal to equation (\ref{eq:eq4}) and equation (\ref{eq:eq5}), and
solve for $\frac{dn_{m}}{dt}$ and $\frac{dn_{p}}{dt}$.\\ Then, we have
\begin{equation}\label{eq:eq6}
\frac{dn_{m}}{dt}=\frac{3\tau_{p-m}}{M_{m}(GM_{p})^{2/3}}n_{m}^{4/3}
\end{equation}
and
\begin{equation}\label{eq:eq7}
\frac{dn_{p}}{dt}=\frac{3\tau_{p-s}}{M_{p}(GM_{s})^{2/3}}n_{p}^{4/3}.
\end{equation}
When we combine equation (\ref{eq:eq1}), equation (\ref{eq:eq2}), equation (\ref{eq:eq3}), equation (\ref{eq:eq6}), and equation (\ref{eq:eq7}),
we obtain the differential equations that govern the time-evolution of the star-planet-moon system:

\begin{eqnarray}\label{eq:basiceq}
\left\{
\begin{array}{ll}
\frac{dn_{m}}{dt}=-\frac{9}{2}\frac{k_{2p}R^{5}_{p}}{Q_{p}}
           \frac{GM_{m}}{(GM_{p})^{8/3}}n_{m}^{16/3}{\rm sgn}(\Omega_{p}-n_{m})\\
        \frac{dn_{p}}{dt}=-\frac{9}{2}\frac{k_{2p}R^{5}_{p}}{Q_{p}}
            \frac{1}{(GM_{p})(GM_{s})^{2/3}}n_{p}^{16/3}{\rm sgn}(\Omega_{p}-n_{p})\\
\frac{d\Omega_{p}}{dt}=-\frac{3}{2}\frac{k_{2p}R^{3}_{p}}{Q_{p}}
        \frac{(GM_{m})^2}{\alpha(GM_{p})^{3}}n_{m}^{4}{\rm sgn}(\Omega_{p}-n_{m})\\
            \qquad\qquad-\frac{3}{2}\frac{k_{2p}R^{3}_{p}}{Q_{p}}
        \frac{1}{\alpha(GM_{p})}n_{p}^{4}{\rm sgn}(\Omega_{p}-n_{p}).
\end{array}
\right.
\end{eqnarray}

The solutions to these differential equations are
\begin{subequations}
\begin{eqnarray}
&&n_{m}(t)=\left(\frac{39}{2}\frac{k_{2p}R_{p}^{5}}{Q_{p}}\frac{GM_{m}}{(GM_{p})^{8/3}}\nonumber
             \:t \:{\rm sgn}(\Omega_{p}-n_{m})\right.\\
            && \left.\qquad\qquad\qquad\qquad\qquad\qquad\qquad+n_{m}^{-13/3}(t=0)\right)^{-3/13}\\
&&n_{p}(t)=\left(\frac{39}{2}\frac{k_{2p}R_{p}^{5}}{Q_{p}}\frac{1}{(GM_{p})(GM_{s})^{2/3}}\nonumber
             \:t \:{\rm sgn}(\Omega_{p}-n_{p})\right.\\
             &&\left.\qquad\qquad\qquad\qquad\qquad\qquad\qquad+n_{p}^{-13/3}(0)\right)^{-3/13}\\
&&\Omega_{p}(t)=-\frac{1}{ \alpha R_{p}^{2}}\left\{\frac{GM_{m}}{(GM_{p})^{1/3}}\left(n_{m}^{-1/3}(t)-n_{m}^{-1/3}(0)\right)\right.\nonumber\\
          &&\left.\qquad\qquad+(GM_{s})^{2/3}\left(n_{p}^{-1/3}(t)-n_{p}^{-1/3}(0)\right)\right\}\nonumber\\
          &&\qquad\qquad+\Omega_{p}(0)\label{eq:sol}.
\end{eqnarray}
\end{subequations}


These solutions are only valid if each of sgn$(\Omega_{p}-n_{m})$ and sgn$(\Omega_{p}-n_{p})$ are constant in time.
Also, these solutions are only valid when the planet's rotation is not tidally synchronous with either the star or the moon, i.e. $\Omega_{p}-n_{m}\neq0$ and $\Omega_{p}-n_{p}\neq0$.
When the planet's rotation is synchronized, we must use an alternative approach.

When synchronization has occurred, i.e. when $\Omega_{p}=n_{m}$ or $\Omega_{p}=n_{p}$, we follow the evolution of the system using conservation of angular momentum:
\begin{equation}\label{eq:ang}
\frac{M_{m}(GM_{p})^{2/3}}{n_{m}^{1/3}(t)}
+\alpha R_{p}^{2}M_{p}\Omega_{p}(t)+\frac{M_{p}(GM_{s})^{2/3}}{n_{p}^{1/3}(t)}
=L_{0},
\end{equation}
where $L_{0}=\frac{M_{m}(GM_{p})^{2/3}}{n_{m}^{1/3}(0)}
+\alpha R_{p}^{2}M_{p}\Omega_{p}(0)+\frac{M_{p}(GM_{s})^{2/3}}{n_{p}^{1/3}(0)}$ is the initial angular momentum.

By our assumption 2, the total angular momentum is the sum of the moon's orbital angular momentum, which is the first term, the planet's rotational angular momentum, which is the second term, and orbital angular momentum, which is the third term.



When the planet is not tidally locked with either the star or the moon, these three equations are valid:
\begin{subequations}\label{eq:sol2}
\begin{eqnarray}
&&n_{m}(t)=\left(\frac{39}{2}\frac{k_{2p}R_{p}^{5}}{Q_{p}}\frac{GM_{m}}{(GM_{p})^{8/3}}\:t \:{\rm sgn}(\Omega_{p}-n_{m})\right.\nonumber\\
             &&\left.\qquad\qquad\qquad\qquad\qquad\qquad+n_{m}^{-13/3}(0)\right)^{-3/13}\label{eq:sol2a}\\
&&n_{p}(t)=\left(\frac{39}{2}\frac{k_{2p}R_{p}^{5}}{Q_{p}}\frac{1}{(GM_{p})(GM_{s})^{2/3}}\:t \:{\rm sgn}(\Omega_{p}-n_{p})\right.\nonumber\\
             &&\left.\qquad\qquad\qquad\qquad\qquad\qquad+n_{p}^{-13/3}(0)\right)^{-3/13}\label{eq:sol2b}\\
&&\frac{(GM_{m})(GM_{p})^{2/3}}{n_{m}^{1/3}(t)}+\alpha R_{p}^{2}(GM_{p})\Omega_{p}(t)\nonumber\\
            &&\qquad\qquad\qquad\qquad+\frac{(GM_{p})(GM_{s})^{2/3}}{n_{p}^{1/3}(t)}=GL_{0}\label{eq:sol2c}.
\end{eqnarray}
\end{subequations}

Even though equation (\ref{eq:sol}) and equation (\ref{eq:sol2c}) are equivalent,
equation (\ref{eq:sol2c}) is valid when the planet's rotation is tidally locked to the moon because equation (\ref{eq:sol2c}) is derived from the conservation of angular momentum.


When the planet is tidally locked with the moon, i.e. $n_{m}=\Omega_{p}$, equation (\ref{eq:sol2a}) is not valid. Hence, in that case,
\begin{subequations}\label{eq:sol2`}
\begin{eqnarray}
&&n_{p}(t)=\left(\frac{39}{2}\frac{k_{2p}R_{p}^{5}}{Q_{p}}\frac{1}{(GM_{p})(GM_{s})^{2/3}}\:t \:{\rm sgn}(n_{m}-n_{p})\right.\nonumber\\
             &&\left.\qquad\qquad\qquad\qquad\qquad\qquad+n_{p}^{-13/3}(0)\right)^{-3/13}\label{eq:sol2`a}\\
&&\frac{(GM_{m})(GM_{p})^{2/3}}{n_{m}^{1/3}(t)}+\alpha R_{p}^{2}(GM_{p})n_{m}(t)\nonumber\\
            &&\qquad\qquad\qquad\qquad+\frac{(GM_{p})(GM_{s})^{2/3}}{n_{p}^{1/3}(t)}=GL_{0}\label{eq:sol2`b}.
\end{eqnarray}
\end{subequations}

When the planet is tidally locked with the star, i.e. $n_{p}=\Omega_{p}$, equation (\ref{eq:sol2b}) is not valid. Hence, in that case,
\begin{subequations}\label{eq:sol2``}
\begin{eqnarray}
&&n_{m}(t)=\left(\frac{39}{2}\frac{k_{2p}R_{p}^{5}}{Q_{p}}\frac{GM_{m}}{(GM_{p})^{8/3}}\:t \:{\rm sgn}(n_{p}-n_{m})\right.\nonumber\\
             &&\left.\qquad\qquad\qquad\qquad\qquad\quad+n_{m}^{-13/3}(0)\right)^{-3/13}\label{eq:sol2``a}\\
&&\frac{(GM_{m})(GM_{p})^{2/3}}{n_{m}^{1/3}(t)}+\alpha R_{p}^{2}(GM_{p})n_{p}(t)\nonumber\\
            &&\qquad\qquad\qquad\quad+\frac{(GM_{p})(GM_{s})^{2/3}}{n_{p}^{1/3}(t)}=GL_{0}\label{eq:sol2``b}.
\end{eqnarray}
\end{subequations}
\section{Numerical Solutions}
We first explore the implications of equations (\ref{eq:sol2}), (\ref{eq:sol2`}), and (\ref{eq:sol2``}) numerically.
In these simulations, we start with $\Omega_{p}(0)>n_{m}(0)>n_{p}(0)$.
Physically, this condition implies that one planet year is longer than one planet day and that the orbital period of the moon is between them.
A typical example is our Sun-Earth-Moon system.
The shapes of the resulting graphs of $\Omega_{p}$, $n_{m}$, and $n_{p}$ as a function of time depend on the torques due to the planet and moon, but also on the orbital angular velocities of the planet and moon.
If both the orbital angular velocities of the moon and planet are slower than the spin angular velocity of the planet, then
both the torques due to the moon and those due to the star brake the rotation of the planet.
If the orbital angular velocity of the moon is faster than the spin angular velocity of the planet, then
the spin angular velocity of the planet may increase or decrease, depending on the relative magnitude of the torque due to the moon and star.

One important lunar escape condition for the calculation is the critical semimajor axis; this is the outermost stable orbit for the moon.
\citet{2002ApJ...575.1087B} stated that the critical semimajor axis, $a_{crit}$, is
\begin{equation}\label{eq:acrit}
a_{crit}=f R_{H},
\end{equation}
where $f$ is a constant and $R_{H}$ is the radius of the Hill's sphere \citep{dePater.Lissauer}

\begin{equation}\label{eq:hills}
R_{H}=a_{p} \left(\frac{M_{p}}{3M_{s}}\right)^{1/3},
\end{equation}
where $a_{p}$ is the semimajor axis of the planet.
Orbits outside $a_{crit}$ are not stable.
In this paper, we follow the orbits using angular velocity instead of semimajor axis.
Considering $n_{crit}$ such that
\begin{equation}
n^{2}_{crit}a^{3}_{crit}=GM_{p},
\end{equation}
then, using (\ref{eq:acrit}) and (\ref{eq:hills}), and Kepler's Third Law for the planet, $n^{2}_{p}a^{3}_{p}=GM_{s}$,
we calculate $n_{crit}$ as a function of $f$:
\begin{equation}\label{eq:ncrit}
n_{crit}(t)=\left(\frac{3}{f^3}\right)^{1/2}n_{p}(t).
\end{equation}
The value of $f$ is not well-determined.
\citet{2002ApJ...575.1087B} used $f=0.36$.  \citet{2006MNRAS.373.1227D} suggested $f=0.49$.
In this paper, we use $f=0.36$ for numerical calculations because it is the most conservative estimate for the moon to remain bound.

This is the critical mean motion that is the lowest stable angular velocity for the moon.
From section \ref{TIS} to section \ref{TIIIS}, we enforce that
\begin{equation}
n_{crit}(t)<n_{m}(t).
\end{equation}
This means that the moon has a stable orbit.

In the resulting numerical integrations, we found three classes of stable outcomes.
We call the three stable outcomes :
\begin{itemize}
    \item Type I(Fig.\ref{fig:TyIC1} and \ref{fig:TIC2})
        \begin{itemize}
            \item planet-moon become synchronous
        \end{itemize}
    \item Type II(Fig.\ref{fig:T123})
        \begin{itemize}
           \item planet-star become synchronous first, then planet-moon become synchronous later
        \end{itemize}
    \item Type III(Fig.\ref{fig:TypeIII})
        \begin{itemize}
           \item  planet-moon never synchronous.
        \end{itemize}
\end{itemize}

For Type I, there are two subcases.

In each of the three stable outcomes, the first part is common.
Initially, $\Omega_{p}(0)>n_{m}(0)>n_{p}(0)$.
Since the orbital angular velocities of the moon and the planet are slower than the spin angular velocity of the planet,
the torques due to the moon and the star brake the rotation of the planet.
This continues until the spin angular velocity of the planet is equal to the angular velocity of the moon.
We call this time $T1$.
From the beginning to $T1$, the planet loses rotational angular momentum to the orbital motions of the moon and the planet.
By gaining angular momentum, the orbital motions of the moon and the planet slow down and their semimajor axes increase.

After $T1$, each type has its own characteristics.
Another feature that Type I, II, and III have in common is that the planet's angular velocity, $n_{p}(t)$, always decreases due to solar tides.
This indicates that the orbital angular momentum of the planet always increases.

In Type I, if the tidal locking starts at $T1$, then the system is Type I Case 1 (Fig.\ref{fig:TyIC1}).
The system is Type I Case 2 if the tidal locking starts after $T1$ (Fig.\ref{fig:TIC2}).
\begin{itemize}
    \item Type I
    \begin{itemize}
        \item Case1 (Fig.\ref{fig:TyIC1})
        \begin{figure}[hbt]
            \begin{center}
            \includegraphics[width=100mm]{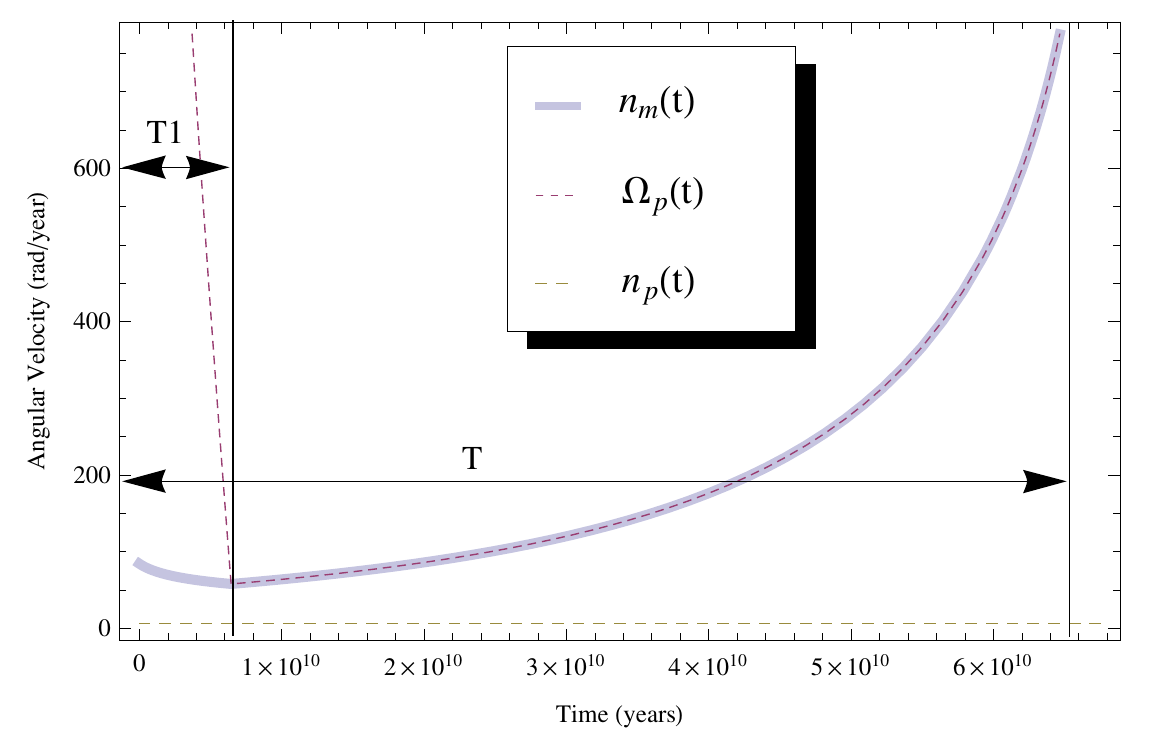}
            \end{center}
            \caption{Here we graph the time evolution of $\Omega_{p}$, $n_{p}$, and $n_{m}$ for Type I Case 1.
                    We use the present data of our Sun-Earth-Moon system for the initial condition, i.e. $n_{m}(0)$=84 rad/year, $\Omega_{p}(0)=730 \pi$ rad/year, and $n_{p}(0)=2 \pi$ rad/year. We take $k_{2p}$ and $Q_{p}$ for Earth to be 0.299 and 12 respectively (\citet{SolarSystemDynamics}, pg166). The black vertical line on the right corresponds to the lunar orbital frequency at which the moon is orbiting at the planetary radius, i.e.when it crashes into the Earth and is destroyed.}
            \label{fig:TyIC1}
            \end{figure}
        \begin{itemize}
        \item
         In our Type I Case 1 star-planet-moon system, the torque on the planet due to the moon is greater than that due to the star at $t=T1$. At $T1$, the planet and the moon then assume a synchronized state with $\Omega_{p}=n_{m}$.
        Once they reach this synchronized state, they will stay in this state until the end for Type I Case 1.
        Since the tidal torque on the planet due to the moon is greater than that due to the star,
        the moon's orbital velocity, $n_{m}$, and the planet's spin angular velocity, $\Omega_{p}$, are kept equal.
        In this synchronized state, only the orbital motion of the moon loses angular momentum; the planet's orbital and spin motion gain angular momentum.
        \end{itemize}
        \item Case2(Fig.\ref{fig:TIC2})
        \begin{figure}[hbt]
            \begin{center}
            \includegraphics[width=100mm]{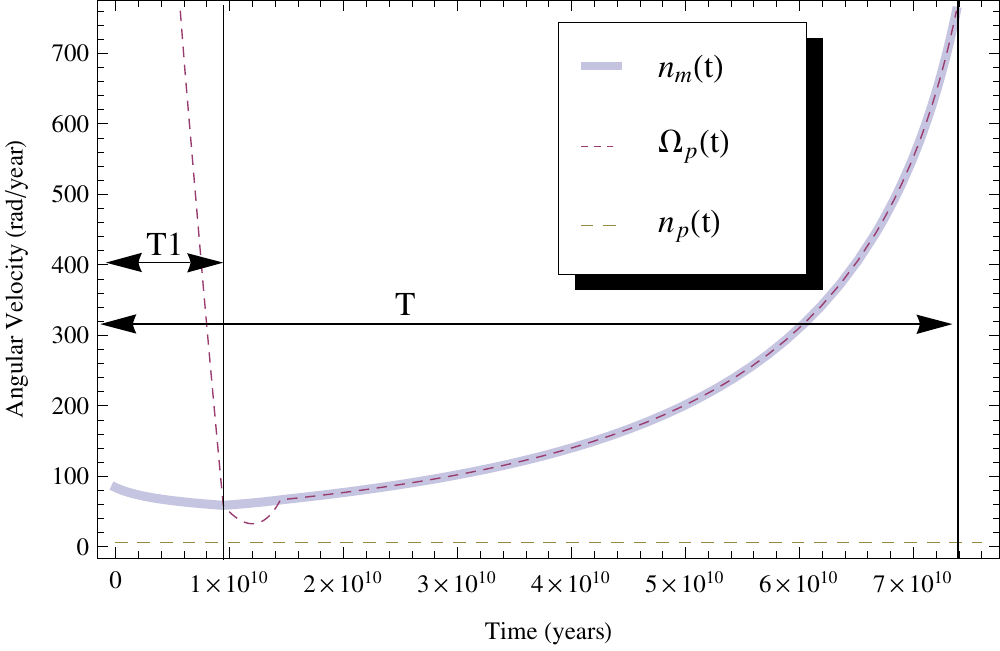}
            \end{center}
            \caption{This graph is for Type I Case 2.
                    We use the same conditions as in Fig.\ref{fig:TyIC1} except the planet's mass is 1.2 $M_{\oplus}$. Note the "notch" in $\Omega_{p}$ just after $T1$; it is what differentiate Case2 from Case1.}
            \label{fig:TIC2}
            \end{figure}
        \begin{itemize}
        \item In Type I Case 2, the moon's tidal torque on the planet is slightly smaller than the stellar torque at $T1$, but
            the planet's rotation never becomes tidally locked to the star.
            There is a brief period when $\Omega_{p}$ is between $n_{m}$ and $n_{p}$, tidally locked to neither the star nor the planet.
            At $t=T1$, the planet and the moon cannot reach the synchronized state because the torque due to the moon is smaller than that due to the star.
            The planet's spin keeps decreasing for a while.
            In this period, the moon's orbital motion and the planet's spin motions lose angular momentum, and the planet's orbital angular momentum increases because of the decreasing semimajor axis of the moon.
            As the moon's orbital angular velocity, $n_{m}(t)$, increases, so does the tidal torque due to the moon.
            Shortly thereafter, the torque due to the moon overcomes the torque due to the star.
            The planet's spin angular velocity, $\Omega_{p}(t)$, starts to increase.
            Then, the planet and the moon reach the synchronized state.
            Once synchronous, the moon's orbital motion loses angular momentum, and the planet's orbital and spin motions increase angular momentum.
            This case is distinct from Case 1 in that there is a period when
            the moon is migrating inward, but is not synchronized with the planet's spin.
        \end{itemize}
    \end{itemize}
    \item Type II(Fig.\ref{fig:T123})
            \begin{figure}[htb]
            \begin{center}
            \includegraphics[width=100mm]{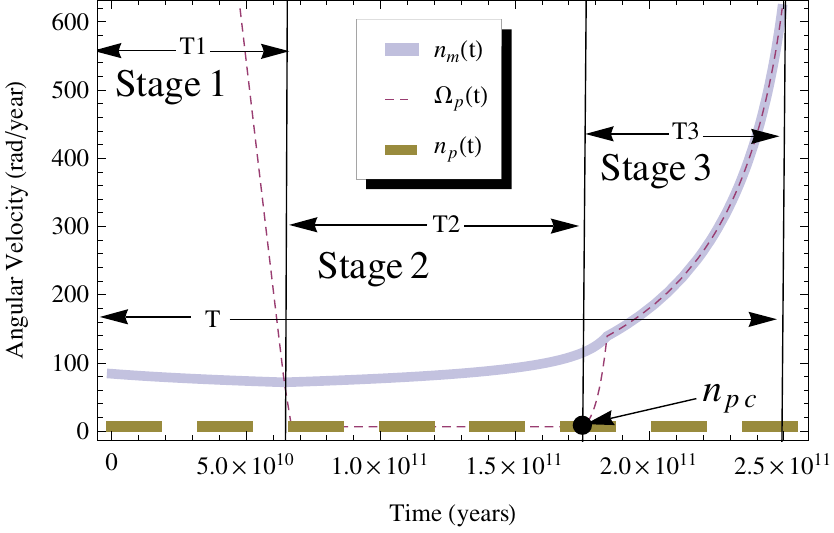}
            \end{center}
            \caption{This graph is for Type II.
                    We use the same conditions as in Fig.\ref{fig:TyIC1} except the planet's mass is 4 $M_{\oplus}$.
                    Here the planetary spin becomes synchronous with its orbit after $T1$ for a time $T2$. But spins up to become synchronous with the moon thereafter.}
            \label{fig:T123}
            \end{figure}
    \begin{itemize}
    \item For Type II, at $t=T1$, the tidal torque due to the star is greater than that due to the moon, which forces the planet's rotation to continue to slow down until it becomes synchronized to the star ($\Omega_{p}=n_{p}$).

             In Stage 2, then, the planet and star remain in a synchronized state because the torque due to the moon does not overcome the torque due to the star.
             Until the planet and star reach a synchronized state, the moon's orbital motion and the planet's spin motion both lose angular momentum.
             At the star-planet synchronized state, only the moon's orbital motion loses angular momentum.

            We can see the difference between Type II and Type III in Stage 3.
            Roughly speaking, if we can see Stage 3, then the system is Type II.
            If Stage 3 is so short that we cannot see it, the system is Type III.
            For Type II, the torque due to the star becomes smaller than the torque due to the moon as the moon spirals inward.
            The planet's rotation becomes tidally locked to the moon, after which only the moon's orbital motion loses angular momentum.
    \end{itemize}
    \item Type III(Fig.\ref{fig:TypeIII})
    \begin{itemize}
    \item In this case, the tidal torque due to the star is always greater than that due to the moon.
            The amount of loss or gain in angular momentum for the moon's orbital motion is so small that we can treat the sum of the orbital angular momentum and the spin angular momentum of the planet as a constant.
        In essence, the planet's spin evolves as if the moon does not exist - this corresponds to the \citet{2002ApJ...575.1087B} condition.
    \begin{figure}[hbt]
        \begin{center}
        \includegraphics[width=100mm]{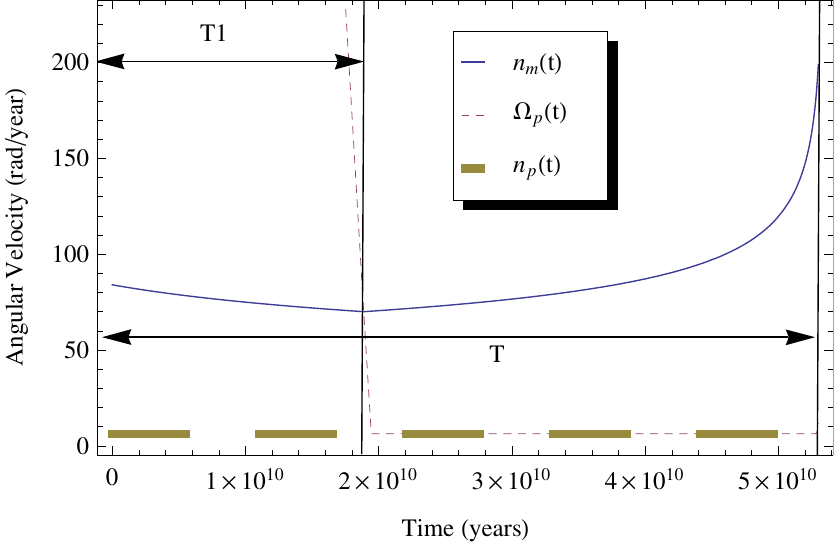}
        \end{center}
        \caption{This graph represents Type III.
                We use the same conditions as in Fig.\ref{fig:TyIC1} except that the moon's mass is set to 0.1 $M_{\rightmoon}$. The planetary rotation becomes synchronous with the star, and never with its moon.}
        \label{fig:TypeIII}
        \end{figure}
    \end{itemize}
\end{itemize}

All $n_{p}$s in Fig.\ref{fig:TyIC1}-Fig.\ref{fig:TypeIII} decrease only a very small amount.
These changes are almost unnoticeable.
However, the differences between the outcomes of tidal evolution in a two body system and a three body system come from these small changes.
Because $n_{p}$ decreases in our model, which does not include the star's tidal response to the planet, our results have a systematic error with respect to the lifetime of moons for close-in planets which experience orbital decay.

\section{Analytical Lifetimes}
In this section, we derive analytical formulae for the total moon lifetime for each of Type I, II, and III.
The total moon lifetime means the time it takes for the moon to either hit the planet or escape.
We will discuss the results here; The full derivations of the formulae are described in the Appendix.
\subsection{Type I Solution}\label{TIS}
As discussed in section 3, Type I has two different cases.
Because we can use the same formula to calculate the lifetime of the moon in each case though, we call both cases Type I.
By creating a new function, $\widetilde{{n}}_{m}(t)$, which coincided with $n_{m}(t)$ after $T1$,
we found the formula for the lifetime of the moons.

The formula for the lifetime of the moons for Type I, $T$, is:
\begin{eqnarray}\label{eqLFTI}
{\scriptstyle
\left.
\begin{array}{ll}
T=\frac{2}{39}\frac{Q_{p}}{k_{2p}R_{p}^5}(GM_{p})(GM_{s})^{2/3}\\
\qquad\left[\left(\frac{3^{3/4}GL_{0}-4\{(GM_{m})^3(GM_{p})^3\alpha R_{p}^2\}^{1/4}}{3^{3/4}(GM_{p})(GM_{s})^{2/3}}\right)^{13}-\left(\frac{1}{n_{p}(0)}\right)^{13/3}\right].
\end{array}
\right.
}
\end{eqnarray}

\begin{figure}[htb]
\begin{center}
\includegraphics[width=100mm]{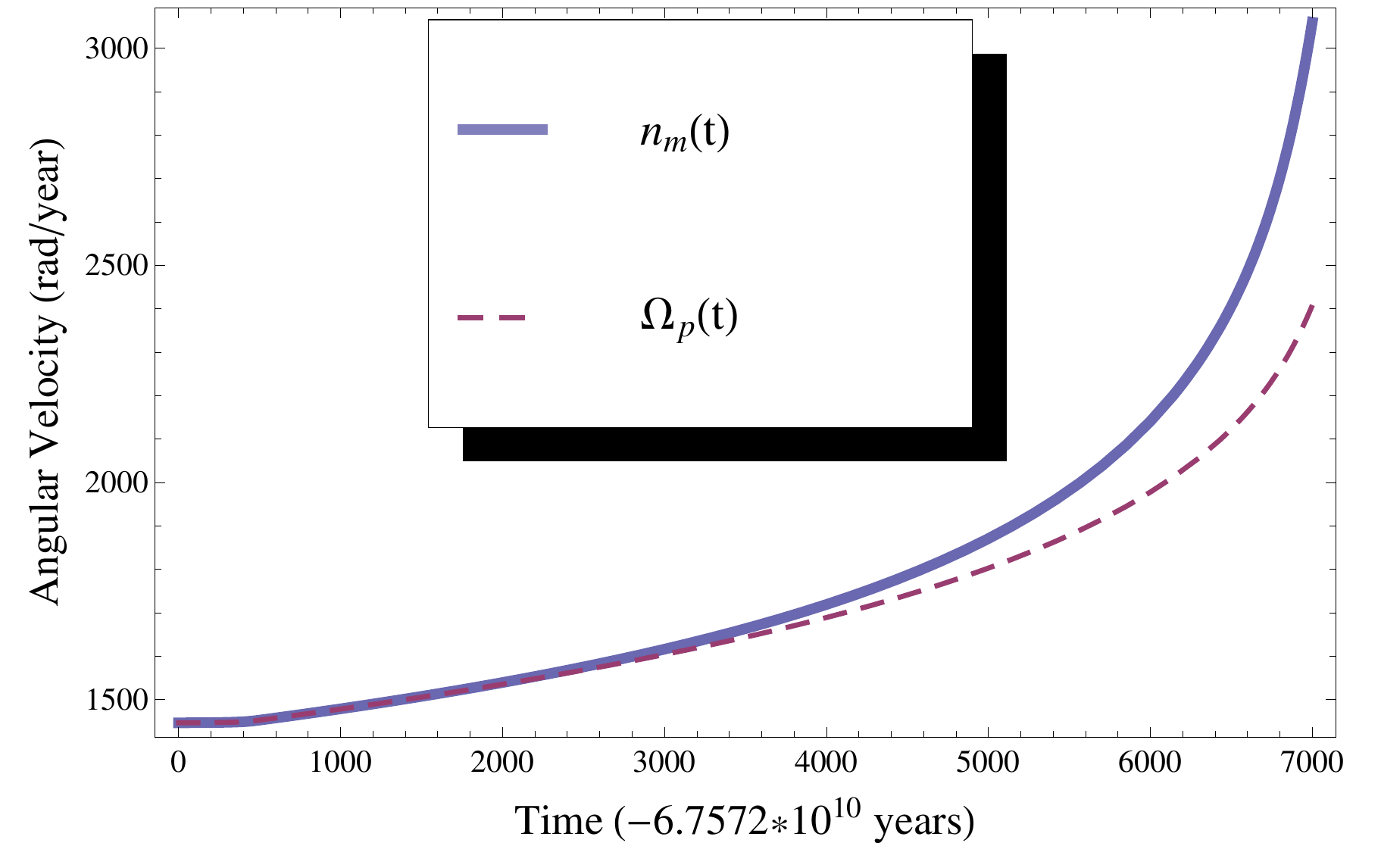}
\end{center}
\caption{This graph is a magnification of the last part of Fig \ref{fig:TyIC1}. At the very end, the synchronized state is over. The orbit of the moon decays inward much faster than before. Because the duration of this final death spiral is so short - 7000 years in a 67.5 Gyr evolution - we neglect it in our analytical formulations.}
\label{fig:CSi}
\end{figure}

After the synchronized state is broken at the very end of the moon's life, the moon has a spiral inward orbit (Fig \ref{fig:CSi}).
We do not explicitly include this period because it is very small compared to $T$.
For Type II, we have the same situation.

Interestingly, a Type I star-planet-moon system experiences all three Counselman states.
From $t=0$ to $T1$, the orbital velocity of the moon, $n_{m}$, decreases.
This indicates that the orbital semimajor axis of the moon increases.
This is Counselman state (ii), except that the moon does not escape from the planet.
At the synchronized state, the orbital velocity of the moon is equal to the spin angular velocity of the planet, i.e. $n_{m}=\Omega_{p}$.
This corresponds to Counselman state (iii).
After the synchronized state, the moon has a brief spiral inward orbit (Fig \ref{fig:CSi}); this is Counselman state (i).


\subsection{Type II Solution}
For Type II, there are three stages (Fig.\ref{fig:T123}).
We calculated the time intervals for $T1$, $T2$, and $T3$, respectively.
By adding them up, we found the lifetime of the moons for Type II.

The formula for the lifetime of the moons for Type II, $T$, is:
\begin{eqnarray}\label{eq:LFTII}
{\scriptstyle
\begin{array}{ll}
T=T1+T2+T3\\
 \quad=2T1+\frac{2}{39}\frac{Q_{p}}{k_{2p}R^5_{p}}
 \left[
 \frac{(GM_{p})^{8/3}}{(GM_{m})}n^{-13/3}_{m}(0)\right.\\
\qquad +\left.\frac{\left(3^{3/4}GL_{0}-4\left\{(GM_{m})^3(GM_{p})^3\alpha R_{p}^2\right\}^{1/4}\right)^{13}}{3^{39/4}(GM_p)^{12}(GM_{s})^{8}}\right.\\
 \qquad\left.-\frac{(GL_0)^{13}}{\left\{(GM_{p})^{1/2}(GM_{m})^{7/6}+(GM_{p})(GM_{s})^{2/3}\right\}^{12}}\right].
\end{array}
}
\end{eqnarray}

We could not find the analytical expression for $T1$.
However, we can calculate $T1$ by solving the following systems of equations numerically for $t$
\begin{eqnarray}\label{eq:T1}
\left\{
\begin{array}{ll}
n_{m}(t)=\left(\frac{39}{2}\frac{k_{2p}R_{p}^{5}}{Q_{p}}\frac{GM_{m}}{(GM_{p})^{8/3}}
             \:t+n_{m}^{-13/3}(0)\right)^{-3/13}\\
n_{p}(t)=\left(\frac{39}{2}\frac{k_{2p}R_{p}^{5}}{Q_{p}}\frac{1}{(GM_{p})(GM_{s})^{2/3}}
             \:t+n_{p}^{-13/3}(0)\right)^{-3/13}\\
\frac{M_{m}(GM_{p})^{2/3}}{n_{m}^{1/3}(t)}+\alpha R_{p}^{2}M_{p}n_{m}(t)\\
\qquad\qquad\qquad\qquad\qquad+\frac{M_{p}(GM_{s})^{2/3}}{n_{p}^{1/3}(t)}=L_{0}.
\end{array}
\right.
\end{eqnarray}

After Stage 3, there is a brief Stage 4 wherein the moon makes its final death spiral into the planet's cloud tops.
At Stage 4, $n_{m}(t)\neq\Omega_{p}(t)$  - actually, $n_{m}(t)>\Omega_{p}(t)$.
Since $T4$ is very small compared to $T1$, $T2$, and $T3$, we do not explicitly include $T4$ in our calculation.

Type II has all three Counselman states, like Type I, plus one extra state.
Stage 1 corresponds to Counselman state (ii) except that the moon does not escape from the planet.
Stage 2 is the extra state.
At this stage, the planet and the star are tidally locked. Because \citet{1973ApJ...180..307C} considered a planet-satellite system, the planet could not be tidally locked with the star, hence Stage 2 has no corresponding Counselman state. Stage 3 corresponds to Counselman state (iii). The planet and the moon reach a synchronized state. At Stage 4, the moon has a spiral inward orbit. This is Counselman state (i).

\subsection{Type III Solution}\label{TIIIS}
\begin{figure}[hbt]
\begin{center}
\includegraphics[width=100mm]{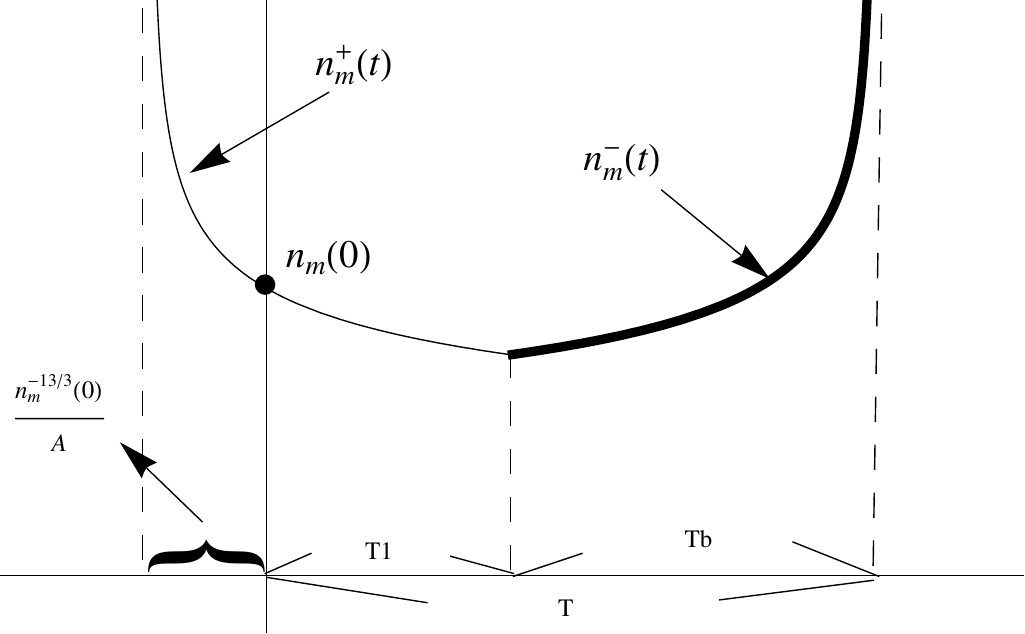}
\end{center}
\caption{This is the graph of $n_{m}(t)$ in Type III.
In this graph, $A=\frac{39}{2}\frac{k_{2p}R_{p}^{5}}{Q_{p}}\frac{GM_{m}}{(GM_{p})^{8/3}}$.}
\label{Fig:ESGTypeIII}
\end{figure}

For Type III, we can calculate the lifetime of the moon using symmetry (Fig \ref{Fig:ESGTypeIII}) as for \citep{2002ApJ...575.1087B}.

The formula for the lifetime of the moons for Type III, $T$, is:
\begin{equation}\label{eq:LFTIII}
T=2T1+\frac{2}{39}\frac{Q_{p}}{k_{2p}R_{p}^{5}}\frac{(GM_{p})^{8/3}}{GM_{m}}{n_{m}^{-13/3}(0)}.
\end{equation}

In general, the system is Type III if the moon is very small compared to the planet.
If we set $(GM_{m})^3(GM_{p})^3\alpha R_{p}^2$ and $(GM_{p})^{1/2}(GM_{m})^{7/6}$ in equation (\ref{eq:LFTII}) equal to zero,
then we get equation (\ref{eq:LFTIII}).

Type III has one Counselman state and one extra state.
The first part, from $t=0$ to $T1$, is Counselman state (ii) except the moon does not escape from the planet.
The second part, $t>T1$, is the extra state which is the same as Stage 2 in Type II.

\subsection{Type IV Solution}\label{TIVS}
\begin{figure}[hbt]
\begin{center}
\includegraphics[width=100mm]{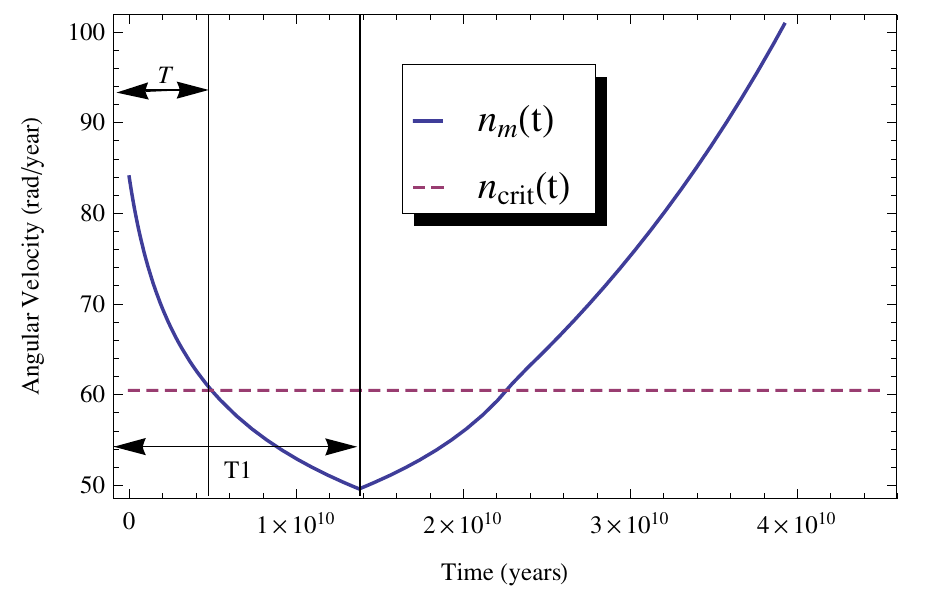}
\end{center}
\caption{This is the graph of $n_{m}(t)$ and $n_{crit}(t)$ for Type IV.
        The dashed line is $n_{crit}(t)$ and the solid line is $n_{m}(t)$.
        At $t=T$, the moon is on the outermost stable orbit.  After that, the orbit becomes unstable.
        Then, the planet loses the moon to interplanetary space. In this case, the lifetime of the moon is $T$. We use the data of the present Sun-Earth-Moon system except with an initial Earth spin angular velocity, $\Omega_{p}(0)$, of $1200\pi$.}
\label{Fig:Type0}
\end{figure}

So far, we assume that the orbit of the moon is always stable.
In this section, we calculate when the orbit of the moon becomes unstable.
This means
\begin{equation}
n_{crit}(t)>n_{m}(t)
\end{equation}
at some $t$ (Fig.\ref{Fig:Type0}).
In this case, we can use equation (\ref{eq:sol2}) with sgn$(\Omega_{p}-n_{m})$=1 and sgn$(\Omega_{p}-n_{p})$=1
because the planet is not tidally locked with either the star or the moon.
To find the time when the orbit of the moon becomes unstable, we set $n_{crit}(t)=n_{m}(t)$.
Then we solve for $t$.
The lifetime of the moon in this case is
\begin{eqnarray}
{\scriptstyle
\left.
\begin{array}{ll}
T=\frac{2}{39}\frac{Q_{p}}{k_{2p}R_{p}^{5}}\left(\frac{(GM_{p})^{8/3}(GM_{s})^{2/3}}{(GM_{m})(GM_{s})^{2/3}-(f^{3}/3)^{13/6}(GM_{p})^{5/3}}\right)\\
\qquad\times\left((f^3/3)^{13/6}n^{-13/3}_{p}(0)-n^{-13/3}_{m}(0)\right).
\end{array}
\right.
}
\end{eqnarray}

Type IV has just one Counselman state: Counselman's state (ii).
\section{Determine the Type of the System}
The expressions in section 4 can be used to calculate the ultimate lifetime of any star-planet-moon system,
provided you know which type of system it is.
In this section, we show how to determine a system's type.
\subsection{Condition for Type I Case 1}
The condition for Type I Case 1 is that the magnitude of the torque due to the moon is greater than the magnitude of the torque due to the star at $t=T1$ (Fig.\ref{fig:TyIC1}),
\begin{equation}
|\tau_{p-m}(T1)|\geq|\tau_{p-s}(T1)|.
\label{TIC1}
\end{equation}

This condition implies
\begin{eqnarray}\label{eq:cdtype1}
T1\leq\frac{2}{39}\frac{Q_{p}}{k_{2p}R_{p}^{5}}\frac{(GM_{p})(GM_{m})^{7/6}(GM_{s})^{2/3}}{(GM_{p})^{1/2}(GM_{s})^{2/3}-(GM_{m})^{7/6}}\nonumber\\
\times\left\{n_{p}^{-13/3}(0)-\left(\frac{GM_{p}}{GM_{m}}\right)^{13/6}n_{m}^{-13/3}(0)\right\}.
\end{eqnarray}
This is the condition for the Type I Case 1.
We can get $T1$ by solving equation (\ref{eq:T1}) numerically.

If the system satisfies equation (\ref{eq:cdtype1}), then we can conclude that it is of Type I Case 1.
If not, the system may be Type I Case 2, Type II, or Type III.

The sign on the right side of equation (\ref{eq:cdtype1}) depends on
\begin{equation}\label{eq:right}
\left\{n_{p}^{-13/3}(0)-\left(\frac{GM_{p}}{GM_{m}}\right)^{13/6}n_{m}^{-13/3}(0)\right\}
\end{equation}
because ${(GM_{p})^{1/2}(GM_{s})^{2/3}-(GM_{m})^{7/6}}>(10^{11/6}-1)(GM_{m})^{7/6}>0$ by our assumption 4.
If equation (\ref{eq:right}) is negative, then the system cannot satisfy equation (\ref{eq:cdtype1}).
Hence, the system is Type I Case 2, Type II, or Type III.
The inequality $\left\{n_{p}^{-13/3}(0)-\left(\frac{GM_{p}}{GM_{m}}\right)^{13/6}n_{m}^{-13/3}(0)\right\}\leq 0$ implies that $|\tau_{p-m}(0)|\leq|\tau_{p-s}(0)|$.
This means that if the initial torque due to the star is greater than the initial torque due to the moon, the system cannot be Type I Case 1.

\subsection{Condition for Type I Case 2}
\begin{figure}[hbt]
\begin{center}
\includegraphics[width=100mm]{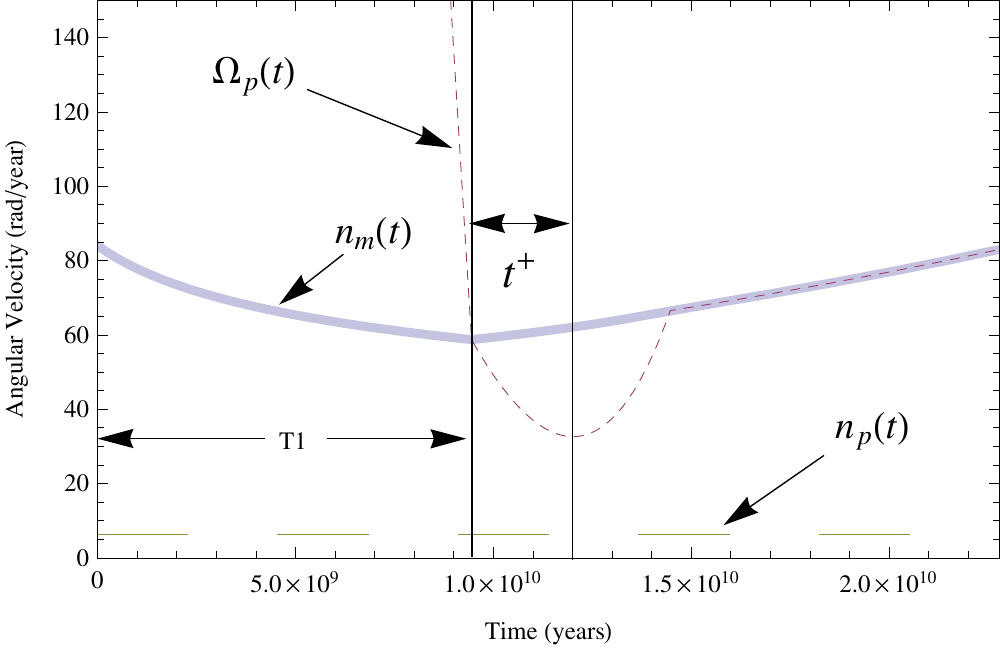}
\end{center}
\caption{This graph is a magnification of Fig \ref{fig:TIC2}.
The rotational rate of the planet, $\Omega_{p}(t)$, decreases until $t=T1+t^{+}$ because the torque due to the star is greater than the torque due to the moon.
At $t=T1+t^{+}$, these two torques are equal.
After that, the torque due to the moon exceeds the torque due to the star.
The rotational rate of the planet, $\Omega_{p}(t)$, starts to increase. Then, the planet and the moon reach a synchronized state.}
\label{fig:TIC2mag}
\end{figure}

Assume $n_{p}(T1)$ and $n_{m}(T1)$ are known.  Let $t^{+}$ be the time from $T1$, when the magnitudes of the two torques are equal (Fig.\ref{fig:TIC2mag}).

The condition for Type I Case 2 is that
\begin{equation}\label{TIC2}
\Omega_{p}(t^{+})\geq n_{p}(t^{+})
\end{equation}
where $t_{*}$ satisfies
\begin{equation}\label{eq:taupmps}
|\tau_{p-m}(t^{+})|=|\tau_{p-s}(t^{+})|.
\end{equation}

This condition implies that
\begin{eqnarray}\label{eq:TIC2S}
a_1 b^{12} X^4-GL_0 X^3+\frac{a_2}{c}b^3\leq0,
\end{eqnarray}
where
\begin{eqnarray}
\small{
\begin{array}{ll}
c=\frac{1}{\alpha R^2_{p} (GM_p)}\nonumber\\
a_1=(GM_p)^{1/2}(GM_m)^{7/78}\nonumber\\
a_2=\frac{1}{(GM_m)^{7/26}}\nonumber\\
b=\left\{(GM_{m})^{7/6}+(GM_{p})^{1/2}(GM_{s})^{2/3}\right\}^{1/13}\nonumber\\
X=\left\{\left(\frac{GM_{p}}{GM_{m}}\right)^{13/6}n_{m}^{-13/3}(T1)+\frac{(GM_p)^{1/2}(GM_s)^{2/3}}{(GM_m)^{7/6}}n_{p}^{-13/3}(T1)\right\}^{1/13}.
\end{array}
}
\end{eqnarray}
If the system is not Type I and satisfies equation (\ref{eq:TIC2S}), then it is Type I Case 2.




\subsection{Conditions for Type II and III}

\begin{figure}[hbt]
\begin{center}
\includegraphics[width=100mm]{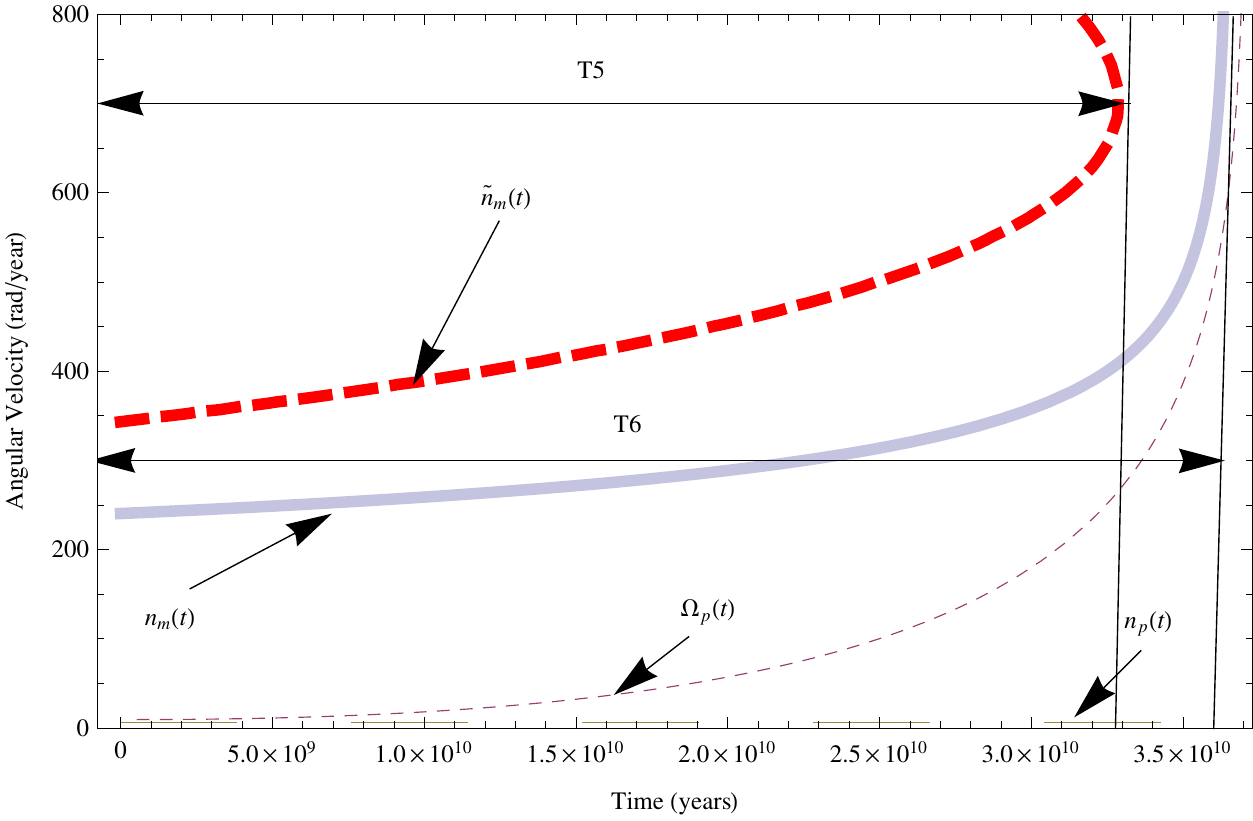}
\end{center}
\caption{
This is a graph of Type III after the synchronized state of the star and planet end.
In this graph, $t=0$ means the time when the synchronized state of the star and planet ends.
We used the present data of our Sun-Earth-Moon system except $M_{p}=18M_{\oplus}$, $n_{m}(0)=240$, $n_{p}(0)=\Omega_{p}(0)=6.28$ for $n_{m}(t)$.
For $\widetilde{n}_{m}(t)$, we used $n_{p}(0)=6.28$, and $\widetilde{n}_{m}(0)=\Omega_{p}(0)=343$.}
\label{fig:ConIII}
\end{figure}

\begin{figure}[hbt]
\begin{center}
\includegraphics[width=100mm]{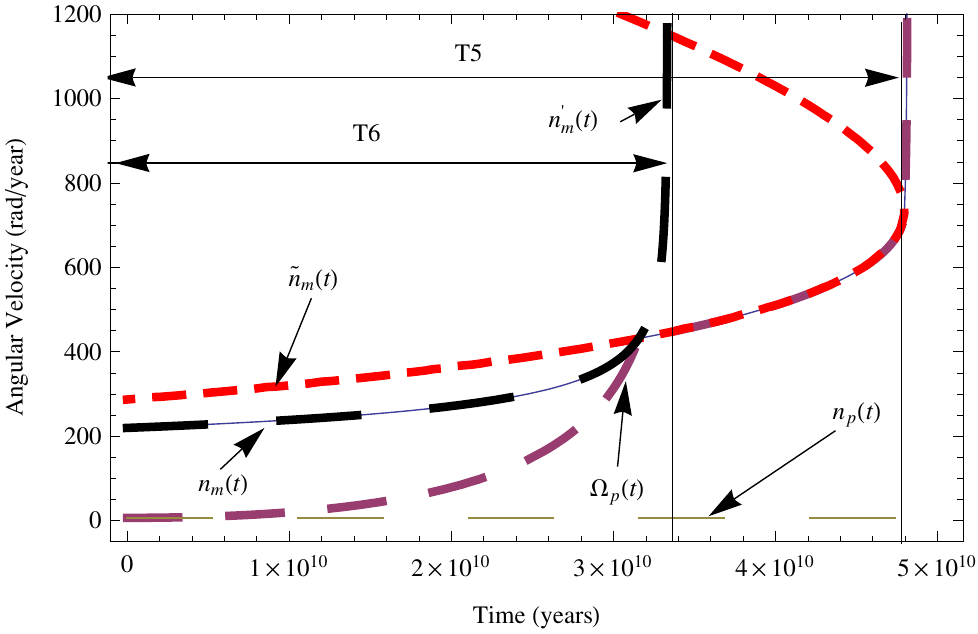}
\end{center}
\caption{This is a graph of Type II at Stage 3.
In this graph, $t=0$ means $t=T1+T2$ in the original Type II graph.
We used the present data of our Sun-Earth-Moon system except $M_{p}=15M_{\oplus}$, $n_{m}(0)=219$, $n_{p}(0)=\Omega_{p}(0)=6.28$.
For $\widetilde{n}_{m}(t)$, we used $n_{p}(0)=6.28$, and $\widetilde{n}_{m}(0)=\Omega_{p}(0)=286$.}
\label{fig:ConII}
\end{figure}

As you can see in Fig.\ref{fig:T123}, the planet and the moon reach a synchronized state at Stage 3 for Type II.
In Fig.\ref{fig:TypeIII}, the synchronized state of the star and planet seems to continue indefinitely.
The planet and star seem to be tidally locked until the end.
Actually, it ends when the moon spirals into the planet, which occurs when $n_{m}(t)$ is large enough.
In Fig.\ref{fig:TypeIII}, the total lifetime of the moon, $T$, is about 53.3 billion years.
The synchronized state of the star and planet ends at about 52.7 billion years.
In Type III, there is a stage that corresponds to Stage 3 for Type II.
However, because that stage is short compared to the total lifetime, it is hard for us to see it.

Fig.\ref{fig:ConIII} is a graph of Type III after the synchronized state of the star and planet ends.
The planet is so big that the spin angular velocity of the planet cannot increase fast enough to catch up to the orbital angular velocity of the moon.
$\widetilde{n}_{m}(t)$ is the hypothetical situation in which the planet and the moon are tidally locked from the beginning.
We introduced $\widetilde{n}_{m}(t)$ in \ref{TIS}.

Fig.\ref{fig:ConII} is a graph of Type II at Stage 3.
The planet is not big enough so that the spin angular velocity of the planet increases fast enough to catch up the orbital angular velocity of the planet.
$n_{m}^{'}(t)$ is the graph of $\left(-\frac{39}{2}\frac{k_{2p}R_{p}^{5}}{Q_{p}}\frac{GM_{m}}{(GM_{p})^{8/3}}\:t+n_{m}^{-13/3}(T1+T2)\right)^{-3/13}$
that is the same equation of $n_{m}(t)$ in Type III.

$T5$ is the maximum range of $\widetilde{n}_{m}(t)$ and $T6$ is the maximum range of $n_{m}(t)$ in Type II and $n_{m}^{'}(t)$ in Type III.

From Fig.\ref{fig:ConIII} and Fig.\ref{fig:ConII}, the condition for Type II is $T5>T6$ and the condition for Type III is $T5\leq T6$.

The condition for Type II, $T5>T6$, implies that
\begin{eqnarray}\label{eq:TII}
{\scriptstyle
\begin{array}{ll}
 \left(3^{3/4}GL_{0}-4\left\{(GM_{m})^3(GM_{p})^3\alpha R_{p}^2\right\}^{1/4}\right)^{13}\times\\
\left((GM_{m})^{7/6}+(GM_{p})^{1/2}(GM_{s})^{2/3}\right)^{12}\\
> 3^{39/4}(GM_{p})^6(GM_{s})^8(GL_{0})^{13}.
\end{array}
}
\end{eqnarray}

Similarly, the condition for Type III, $T5\leq T6$, implies that
\begin{eqnarray}\label{eq:TIII}
{\scriptstyle
\begin{array}{ll}
 \left(3^{3/4}GL_{0}-4\left\{(GM_{m})^3(GM_{p})^3\alpha R_{p}^2\right\}^{1/4}\right)^{13}\times\\
\left((GM_{m})^{7/6}+(GM_{p})^{1/2}(GM_{s})^{2/3}\right)^{12}\\
\leq 3^{39/4}(GM_{p})^6(GM_{s})^8(GL_{0})^{13}.
\end{array}
}
\end{eqnarray}

\subsection{Condition for Type IV}
Looking at the graphs of Type I (Fig.\ref{fig:TyIC1}, Fig.\ref{fig:TIC2}), Type II (Fig.\ref{fig:T123}), and Type III (Fig.\ref{fig:TypeIII}), we know $n_{m}(t)$ has a minimum at $t=T1$.
Hence, the condition for Type IV is
\begin{eqnarray}
{\scriptstyle
\left.
\begin{array}{ll}
T1>T=\frac{2}{39}\frac{Q_{p}}{k_{2p}R_{p}^{5}}\left(\frac{(GM_{p})^{8/3}(GM_{s})^{2/3}}{(GM_{m})(GM_{s})^{2/3}-(f^{3}/3)^{13/6}(GM_{p})^{5/3}}\right)\\
\qquad\times\left((f^3/3)^{13/6}n^{-13/3}_{p}(0)-n^{-13/3}_{m}(0)\right).
\end{array}
\right.
}
\end{eqnarray}

We can find $T1$ by solving the system of equations (\ref{eq:T1}) numerically.

\section{Applications}
In this section, we check the formulae for the lifetime of the moon and the condition for the type of system first in two examples, and then showing an application of our results.

In the first two examples, we use the present data of our Sun-Earth-Moon system.
We take $k_{2p}$ and $Q_{p}$ for Earth to be 0.299 and 12, respectively (\citet{ SolarSystemDynamics}, pg166).
In the first example, we survey a range for the mass of the planet from 0.1 to 25 $M_{\oplus}$.
In the second example, we explore moon masses from 0.01 to 2 $M_{\rightmoon}$.
For Type I, we can plot the graph of the lifetime of the moon easily because every parameter is constant (Eq.\ref{eqLFTI}).
For Type II and III, we need to know the expression for $T1$ to plot the graph (Eq.\ref{eq:LFTII},\ref{eq:LFTIII}).
In order to have $T1$, we must calculate $T1$ numerically.
Then, by using an approximation method, such as best-fitting or interpolation, we can obtain the expression of $T1$.

The fractional error of our analytical formulae is always smaller than $10^{-3}$ when compared to numerical integration and usually much smaller.
The fractional error is defined as the absolute value of the lifetime from simulation minus lifetime from formula divided by lifetime from simulation.

In the first example, $Q_{p}$=12 is not realistic for a high mass planet.
However, we want to show just how the mass of the planet affects the lifetime of the moon.
To get a more realistic result, we should change both $k_{2p}$ and $Q_{p}$ for Jovian planets.
If we did that, we would expect that the lifetime of the moons would be much longer.

In both examples, the lifetime in most of these figures is longer than the typical main sequence lifetimes of solar-type stars.
Over such a timescale additional processes, such as inflation of the stellar radius, and the resulting changes to tidal evolution become important.

\subsection{First Example - Changing The Mass of The Planet}
\begin{figure}[hbt]
\begin{center}
\includegraphics[width=90mm]{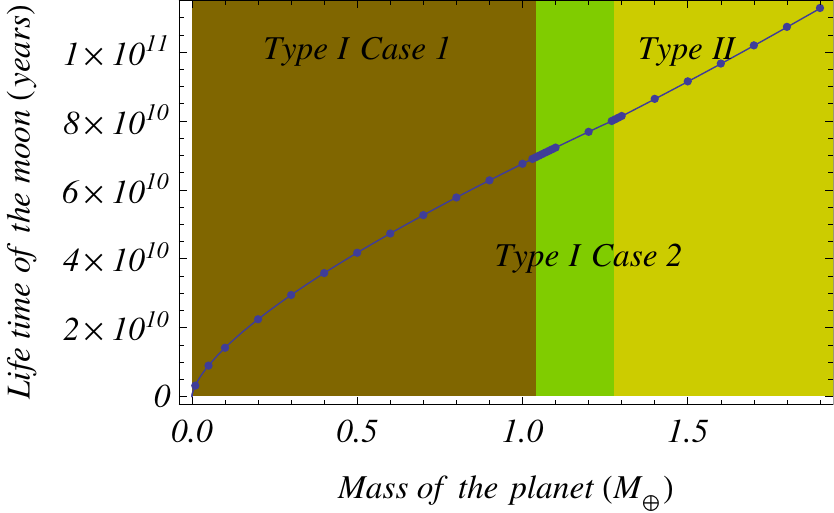}
\end{center}
\caption{This graph shows the lifetime of a hypothetical system with Sun-Earth-Moon parameters with varying planetary mass from 0.1 to 2 $M_{\oplus}$.
Each dot represents the result from numerical solutions, and the curve is generated
by equation (\ref{eqLFTI}) or equation (\ref{eq:LFTII}), depending on the type of system.
The lifetime of the moon increases linearly as the mass of the planet increases in this region.
The borders between Case 1 and Case 2, and Case 2 and Type II are $M_{p}=1.04M_{\oplus}$ and $M_{p}=1.27M_{\oplus}$, respectively. The border between different types depends on the initial conditions and $Q_{p}$.}
\label{fig:MpIColor}
\end{figure}

\begin{figure}[hbt]
\begin{center}
\includegraphics[width=90mm]{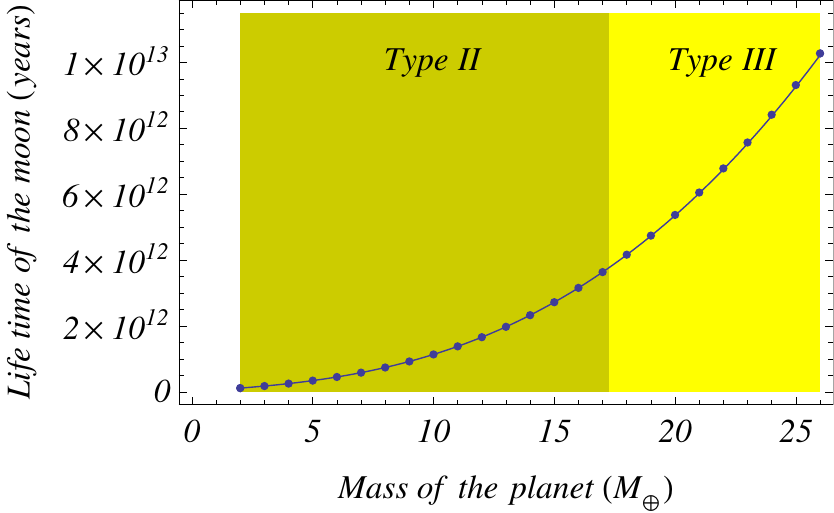}
\end{center}
\caption{Here we show the lifetime of a hypothetical system with Sun-Earth-Moon parameters with varying planetary mass from 2 to 25 $M_{\oplus}$.
Each dot represents the result from numerical solutions.
The curve is generated by equation (\ref{eq:LFTII}), and equation (\ref{eq:LFTIII}).
In this region, the lifetime of the moon increases exponentially as the mass of the planet increases.
The border between Type II and Type II is $M_{p}=17.2M_{\oplus}$. The border between different types depends on the initial conditions and $Q_{p}$.}
\label{fig:MpIIColor}
\end{figure}

Fig.\ref{fig:MpIColor} shows how the lifetime of a 1 $M_{\rightmoon}$ moon varies as the mass of the planet changes from 0.1 to 2 $M_{\oplus}$.
Fig.\ref{fig:MpIIColor} is similar, but with the mass of the planet varying from 2 to 25 $M_{\oplus}$.
The difference between the data from numerical simulations and the graph generated by the formula is very small.
From the graph Fig.\ref{fig:MpIColor}, and Fig.\ref{fig:MpIIColor}, every data point is on the curve.
The transition from one type to another is smooth.

As you can see in Fig.\ref{fig:MpIColor} and Fig.\ref{fig:MpIIColor}, the lifetime of the moon increases as the planet mass increases. The  increased longevity occurs because the effects of the lunar and stellar tides on the planetary spin evolution are reduced as we increase $M_{p}$, i.e. the planet is not braked as easily by each of the these effects as we go to more massive planets. The system therefore continues to evolve with $\Omega_{p}> n_{m}$ for longer, in which the tidal torque on the moon is positive, lengthening the overall evolution.

\subsection{Second Example - Changing The Mass of the Moon}
\begin{figure}[hbt]
\begin{center}
\includegraphics[width=90mm]{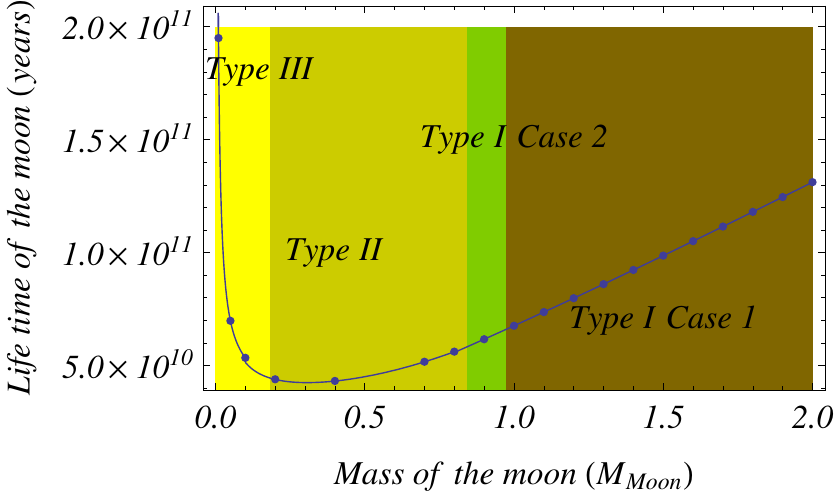}
\end{center}
\caption{This graph shows the lifetime of a hypothetical system with Sun-Earth-Moon parameters with varying moon mass from 0.01 to 2 $M_{\rightmoon}$.
When the mass of the moon is small, the system is Type III.
As the mass of the moon increases, the system becomes Type II, Type I Case 2, and then Type I Case 1.  There is a minimum lifetime of $4.24\times10^{10}$ years at 0.307 $M_{\rightmoon}$.
This minimum arises because the mass of the moon has different effects in Type I and III.
In Type III, the heavier moon has a shorter lifetime due to faster tidal evolution.
While in Type I, the heavier moon has the longer lifetime because the moon has grater orbital angular momentum.
When these effects cancel each other out, there is a minimum value.
The borders between the Type III and Type II, Type II and Case 2, and Case 2 and Case 1 are $M_{m}=0.181M_{\rightmoon}$,
$M_{m}=0.842M_{\rightmoon}$, and $M_{m}=0.972M_{\rightmoon}$, respectively. The border between different types depends on the initial conditions and $Q_{p}$.}
\label{fig:MmColor}
\end{figure}

Fig.\ref{fig:MmColor} shows how a moon's lifetime would change as the moon's mass varies from 0.01 to 2 $M_{\rightmoon}$.
For low-mass moons, the greater the mass of the moon, the more quickly the moon evolves from equation (\ref{eq:basiceq}).
This is why the heavier moon has the shorter lifetime in Type III \citep{2002ApJ...575.1087B}.
Additionally, at Type I Stage 1, a heavy moon evolves faster than a light moon.
Indeed, $T1$ decreases as the mass of the moon increases.

However, once the moon and the planet reach their synchronized state, the planet keeps the heavier moon for a longer time.
When the planet and the moon become tidally locked, this planet-moon system behaves like one object.
As the mass of the moon increases, the moment of inertia of the planet-moon system increases.
Because the star saps angular momentum at a constant rate, the planet-moon system evolves more slowly when the system has the heavier moon.

Between Types III and I, the lifetime of the moon has a minimum value. When the mass of the moon is 0.307 $M_{\rightmoon}$, it has a minimum lifetime of $4.24\times10^{10}$ years.

\begin{figure}[hbt]
            \begin{center}
            \includegraphics[width=90mm]{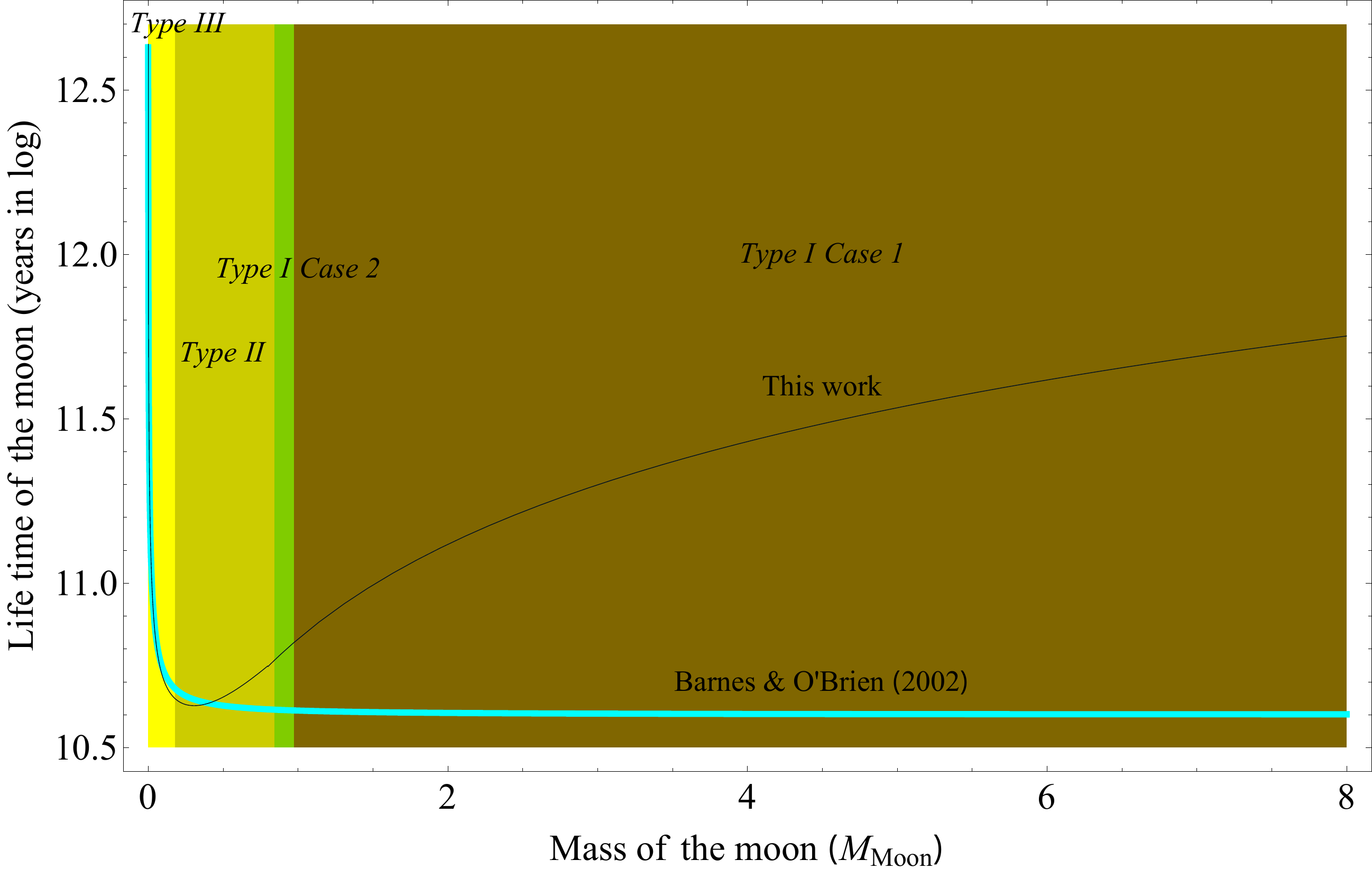}
            \end{center}
            \caption{This graph shows that the lifetime of a hypothetical system
            using Sun-Earth-Moon present parameter with (this work) and without (\cite{2002ApJ...575.1087B}) lunar tides in log scale. The black thin line represents the lifetime of
             the moon including lunar tides. The light blue thick line is the lifetime of
             the moon not including lunar tides. In the Type III region, both results agree very well.
             However, in the Type I region, the necessity of incorporating lunar tides' effect on the planet's rotation, as we have introduced in this paper, becomes clear.
              }
            \label{LT}
            \end{figure}

We know that the tidal effect of the real Moon is important on the Earth. To show the utility of our approach, we calculate the lifetime of hypothetical moons with and without lunar tidal effect (Fig.\ref{LT}). For low masses of the moon there is no difference between these results and previous results of \cite{2002ApJ...575.1087B} because the effect of the lunar tides is small. For the high mass moons, the lifetime of the moon with the lunar tidal effect is significantly longer than that without lunar tidal effect.

\subsection{Third Example - Other Systems}
\begin{figure*}[htb]
\begin{center}
\includegraphics[width=150mm]{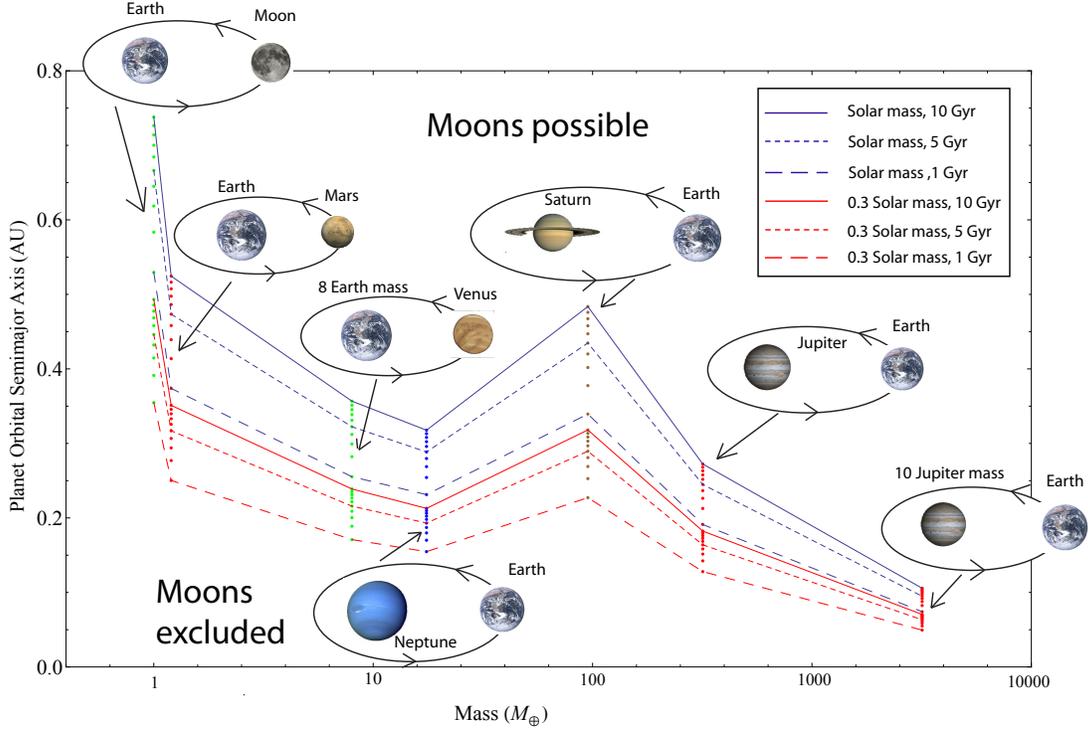}
\end{center}
\caption{This graph shows the location required for the planet/moon system to have 1 Gyr, 2 Gyr, ..., up to 10 Gyr lunar lifetimes.
 The blue lines are for 1.0 $M_{\bigodot}$ stars. The red lines are for 0.3 $M_{\bigodot}$ stars. The pictures show which planet-moon pair the system has. The size of the pictures do not accurately depict the size of the planet and the moon. The top solid lines, both red and blue, are for 10 Gyr lunar lifetimes.  The second and third dashed-lines are for 5 Gyr and 1 Gyr in both red and blue lines. We used the planet-moon synchronized state as the initial condition for each case, i.e $n_{m}(0)=\Omega_{p}(0)$. Each star-planet-moon system has ten dots.
From the bottom to the top, the dots represent 1 Gyr, 2 Gyr, ..., up to 10 Gyr lunar lifetimes. For the Earth, we used a $k_{2p}$ of 0.299, a $Q$ of 12, and the moment inertia constant, $\alpha$ of 0.33. For an 8-Earth-mass-planet, $k_{2p}$ is 0.299, $Q$ is 12, $\alpha$ is 0.33, and the radius of the planet is 1.8 times the Earth radius.  For Neptune, $k_{2p}$ is 0.4, $Q$ is $10^{4}$, and $\alpha$ is 0.23. For Saturn, $k_{2p}$ is 0.35, $Q$ is $1.8\times10^{4}$, and $\alpha$ is 0.21. For Jupiter and the 10 Jupiter mass planet, $k_{2p}$ is 0.5, $Q$ is $10^{5}$, and $\alpha$ is 0.254. We assume that the 10-Jupiter-mass-planet has the same radius as Jupiter. Note that the moon stability lines depend on $Q$. We adopt $Q$=12 for rocky planets, $Q$=$10^{4}\sim 10^{5}$ for ice and gas giants. Increasing $Q$ increases the tidal evolutionary timescales.}
\label{fig:FDM}
\end{figure*}

\begin{figure*}[htb]
\begin{center}
\includegraphics[width=150mm]{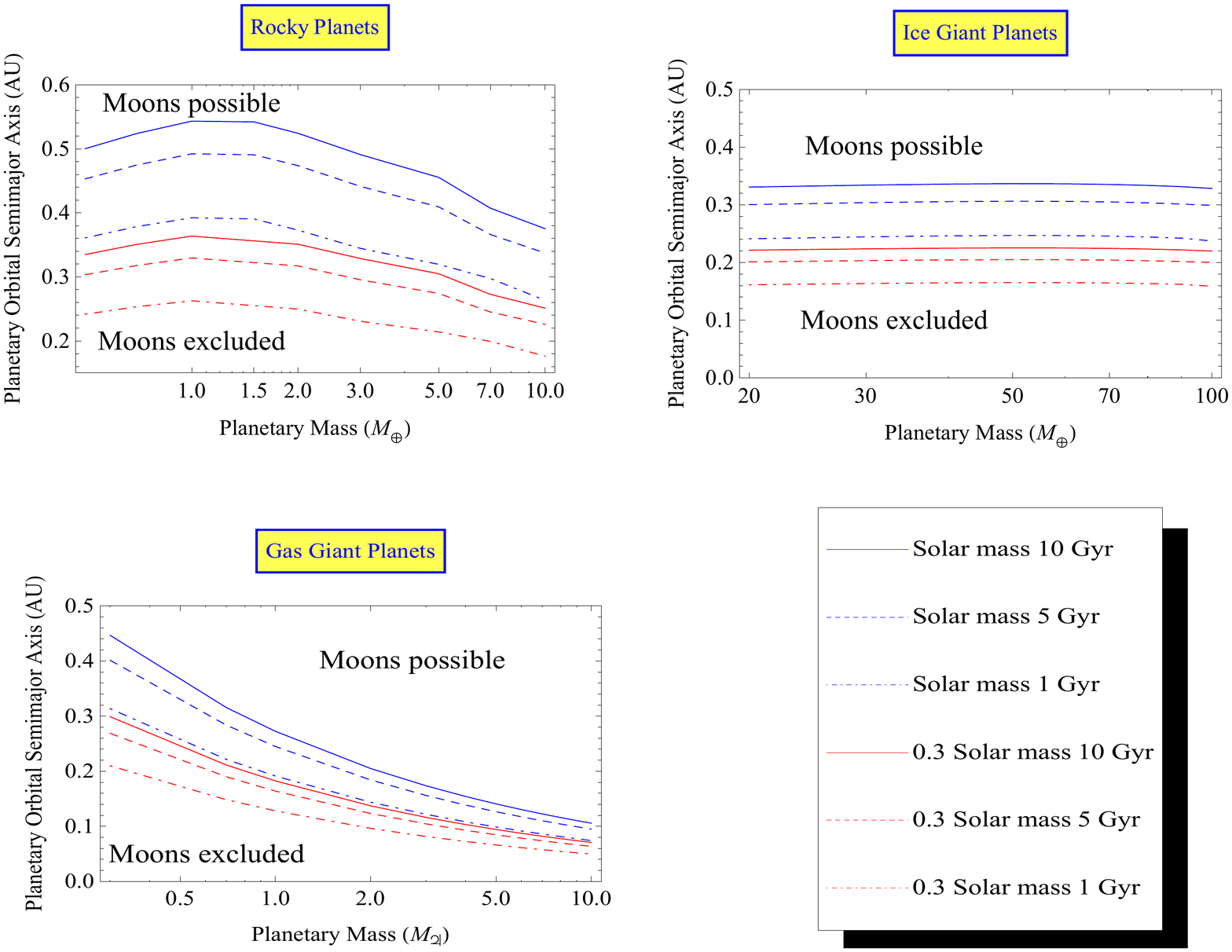}
\end{center}
\caption{The top left graph shows the stability line for rocky planets with our Moon mass moon. We use $k_{2p}$ of 0.299, $\alpha$ of 1/3 and $Q_{p}$ of $100$. We find the planetary radius from \cite{2007ApJ...659.1661F} with the same composition rate of the Earth.
The top right graph shows the stability line for ice giant planets with the Earth mass moon. We use $k_{2p}$ of 0.4, $\alpha$ of 0.23 and $Q_{p}$ of $10^{4}$. We find the planetary radius from \cite{2007ApJ...659.1661F} with the same composition rate of Neptune.
The bottom left graph shows the stability line for gas giant planets with the Earth mass moon. We use $k_{2p}$ of 0.5, $\alpha$ of 0.254 and $Q_{p}$ of $10^{5}$. We assume that the planetary radius is one Jupiter radius.
}
\label{fig:Three}
\end{figure*}

Finally, we apply our results to extrasolar star-planet-moon systems where we expect that the first exomoons will be discovered.
To see the big picture, we chose some typical combinations of the stars, planets, and moons (Fig.\ref{fig:FDM}).
We choose 1.0 $M_{\bigodot}$ and 0.3 $M_{\bigodot}$ stars as the parent stars. For each parent star, we investigate 7 planet/moon systems. For rocky planets, we chose Earth/Moon, Earth/Mars, and 8-$M_{\bigoplus}$-super-earth-planet/Venus systems. For ice giant planets, we choose hypothetical Neptune/Earth system. For gas giant plants, we choose Saturn/Earth, Jupiter/Earth, and 10-$M_{Jup}$-planet/Earth systems.
To see the trend of each type of the planets, we also separate the systems by $Q_{p}$ (Fig.\ref{fig:Three}). We use $Q_{p}=100$, $10^{4}$, and $10^{5}$ for rocky, ice giant, and gas giant planets, respectively.
\begin{figure}[hbt]
\begin{center}
\includegraphics[width=90mm]{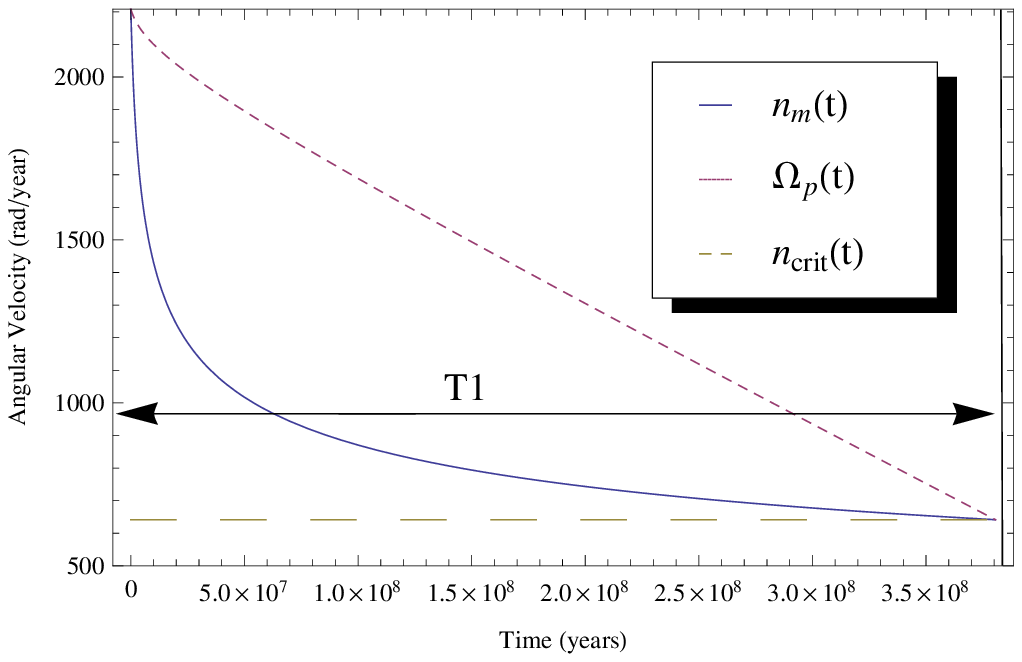}
\end{center}
\caption{This is the first part of a typical graph for the third example. The initial condition is the planet-moon synchronized state, i.e $n_{m}(0)=\Omega_{p}(0)$. The moon almost reached the outer most stable radius, i.e.$n_{m}(T1)\approx n_{crit}(T1)$ and $n_{m}(T1)> n_{crit}(T1)$. The system continues to evolve back inwards towards the planet.}
\label{fig:TEG}
\end{figure}


Fig.\ref{fig:FDM} shows the moon stability lines for 1 to 10 Gyr applied to types of planet/moon systems.
It is worth noting that a star of mass 0.3 $M_{\odot}$ has a main sequence lifetime much longer than that of the Sun.
The lifetimes of the 0.3 $M_{\odot}$ and the Sun are the order of 100 Gyr and 10 Gyr, respectively.
Like the `ice line' with respect to planet formation, we define the `moon stability line' as the location beyond which a moon is stable for the life of the stellar system.
Therefore no such primordial moons can presently exist inside the moon stability line, though moons are possible outside it.
Each point represents a moon stability line.
In each case, we assumed the planet-moon synchronized state, i.e. $n_{m}(0)=\Omega_{p}(0)$ as the initial condition, and the moon almost reached the outer most stable radius, i.e. $n_{m}(T1)\approx n_{crit}(T1)$ and $n_{m}(T1)> n_{crit}(T1)$(Fig.\ref{fig:TEG}).
Because the moon did not reach the outer most stable radius, after that the system continued to evolve back inwards towards the planet.


It is worth noting that the moon stability lines depend on $Q$. In Fig.\ref{fig:FDM}, we adopt $Q$=12 for rocky planets, $Q$=$10^{4}\sim 10^{5}$ for ice and gas giants. Increasing $Q$ increases the tidal evolutionary timescales.

Overall, the moon stability line moves inward for massive planets, for less massive parent stars, and for younger systems.
In other words, moons are more stable when the planet/moon systems are further from the parent star, the planets are heavier, or the parent stars are lighter. This result can be explained by the size of Hill radius. The planet has a larger Hill radius for larger planet mass, smaller stellar mass, or larger planetary semi-major axis. In general, the moon has longer lifetime for the larger Hill radius of the planet.


\paragraph*{Gas giant systems}
For gas giants, the moon stability line moves inward as the mass of the planet increases.
In other words, moons are more stable when gas giants are heavier.
\citet{2002ApJ...575.1087B} studied this type of system.
They concluded that smaller moons are more stable around gas giant planets.
With their high mass ratios, from these results we agree that smaller moons are more stable around heavier gas giants.

\paragraph*{Ice giant systems}
In Fig.\ref{fig:Three} top right, we use $k_{2p}=0.4$, $\alpha=0.23$, $Q_{p}=10^{4}$ which are the parameters of Neptune, and use the Earth as a moon.
In this case, the moon stability line does not move much even the mass of the planet increases.

\paragraph*{Rocky planet systems}
Compared to ice and gas giant planets, rocky planets are small and light.
Therefore, the mass of the moon is one of the factors that moves the moon stability line when compared to the giants.
Look at the Earth/Moon system and the Earth/Mars system in Fig.\ref{fig:FDM}.
The only difference between these systems is the mass of the moon.
As you can see, the moon stability line moves inward as the mass of the moon increases.
Look at the Earth/Mars system and an 8-Earth-mass-planet/Venus system in Fig.\ref{fig:FDM}.
The differences between these systems are the masses of the planet and moon. But the mass ratio between planet and moon is about 10 to 1, which is the maximum ratio of planet and moon for which our formulation is valid, in both cases.
As you can see, the moon stability line moves inward as the mass of the planet and moon increase to a ratio of 10 to 1.
Fig.\ref{fig:Three} top left shows the moon stability lines for rocky planets with our Moon mass moon. We use $k_{2p}=0.299$, $\alpha=1/3$,  and $Q_{p}=100$. The moon stability line moves inward as the mass of the planet increases except below the 1 $M_{\bigoplus}$.

The planet/moon system is preferred to be closer to the parent star to detect an extrasolar moon because we can make many observations of the planet and moon transiting .
However, our result shows that the moon stability line moves inward for a younger system.
If the planet/moon system to the parent star is close, we may find the planet but it's moon has already gone.
If the planet is far from the parent star, it's moon may exist but the planet is hard to detect.
We expect that the semi-major axis of the planet around which the first extramoon of a G-type star is 0.4-0.6 AU because the lifetime of the moon is more than 10 Gyr in most cases and we can observe the transiting planet two to four times in a year.
For M-type star, we expect that the planet/moon system locate 0.2-0.4 AU because the lifetime of the moon is more than 10 Gyr in most cases and we can observe the transiting planet three to six times in a year.

\section{CONCLUSION}
We derive analytical expressions for determining the lifetime of hypothetical moons in star-planet-moon systems.
Our solutions allow us to find the type of system and the lifetime of the moon without the need to numerically solve a system of differential equations.
We first determine whether the moon remains within the planet's outermost stable orbit.
If not, the moon is lost and the system is Type IV.
If the moon remains in orbit, there are three possible outcomes: Types I, II, and III.
In Type I, the planet is tidally locked with the moon.
In Type II, the planet is tidally locked first with the star, and later with the moon.
In Type III, the planet is not tidally locked with the moon.
The type of system depends on characteristics of the star, planet, and moon (mass, radius, Love number $Q_{p}$, etc.)
as well as the initial conditions of the planet and the moon.

Once we determine the system type, we can calculate the lifetime of the moon.
To find the type of system and the lifetime of the moon, we need $T1$, which is the time when the spin angular velocity of the planet is equal to the angular velocity of the moon; See Fig.\ref{fig:TyIC1}, Fig.\ref{fig:T123} and Fig.\ref{fig:TypeIII}.
We should use a numerical method to find $T1$.





Our results are extension of \cite{1973MNRAS.164...21W} and \citet{2002ApJ...575.1087B}.
At the range that they considered, our results agree to their results.
\cite{1973MNRAS.164...21W} considered Type III without critical mean motion.
In this case, the planet will lose its moon only if the moon collides with the planet.
\citet{2002ApJ...575.1087B} considered Type III with critical mean motion.
In this case, the moon may either hit the planet or escape from it.
In both cases, the planet and moon are asynchronous.

\citet{2002ApJ...575.1087B} concluded that the heavier the moon, the shorter the lifetime of the moon.
Because they considered only systems of Type III, this result agree to our result (Fig.\ref{fig:MmColor}). On the other hand, the heavier the moon, the longer the lifetime of the moon for Type I and II.

Our Moon stabilizes Earth obliquity - a key reason for the development of life on Earth \citep{2000rewc.book.....W}. Stable obliquity in its star's habitable zone may be necessary for a planet to support life. An extrasolar moon of sufficient mass could stabilize the obliquity of an Earth-size extrasolar planet. However even if a planet has a relatively large moon like the Earth does, the planetary obliquity may not be stable in some cases such as the moon is far from the planet, the planet is close to the star, there is a Jupiter-size-planet close enough to the planet, etc \citep{2012Icar..217...77L}. On the other hand, Mars has relatively small satellites, and its obliquity changes chaotically, fluctuating on a 100,000-year timescale \citep{1993Natur.361..608L}. Having a relatively large moon is not enough in and of itself to provide a sufficient condition for an extrasolar planet to stabilize its obliquity meaning support life.  Hence, our results give a condition needed to support life on a planet in the habitable zone.

Suppose we find a Jupiter-size planet in the habitable zone.
This planet may have an Earth-sized moon.
If the lifetime of that extrasolar moon is equal to or greater than the age of Earth, then the moon may support life.
Hence, our results gives a condition needed for potentially habitable moons.


In the third example, we show the moon stability lines for 1 to 10 Gyr applied to types of planet/moon systems.
We define the `moon stability line' to be the location beyond which a moon is stable for the life of the stellar system.
In general, the moon stability line moves inward for more massive planet, for a less massive parent star, and for younger systems.
In other words, moons are more stable when the planet/moon systems are further from the parent star, the planets are heavier, or the parent stars are lighter. We expect that the semi-major axis of the planet for the first extramoon of a G-type star will be 0.4-0.6 AU and for an M-type star 0.2-0.4 AU.

This lays the ground work for the tidal evolution of a star-planet-moon system and makes it possible to classify star-planet-moon systems and providing useful estimates of the lifetime of a moon.

In some cases, we may not necessarily be able to accurately predict the long-term survival of the moon.
The value of $Q_{p}$, the specific dissipation function of the planet, is assumed to be constant in time.
However, $Q_{p}$ is not known theoretically, and may depend on the planetary internal structure.
For the sake of simplicity, we considered a star-planet-moon system with a single planet and a single moon.
But we do not consider any interactions between the star and the moon.
This deficiency may be addressed in future work.
Gravitational perturbations caused by other planets or moons may be significant.
For close-in planets, the stellar gravitational perturbations of the moon's orbit are important \citep{2009ApJ...704.1341C}.
In these situations, our method may not predict the lifetime of the moon accurately. Despite its shortcomings, our approach provides an important step toward understanding the tidal evolution and longevity of extrasolar moons, and will form both a basis for future theoretical investigations and direction for future searches to detect extrasolar moons.


\begin{figure}[hbt]
\begin{center}
\includegraphics[width=100mm]{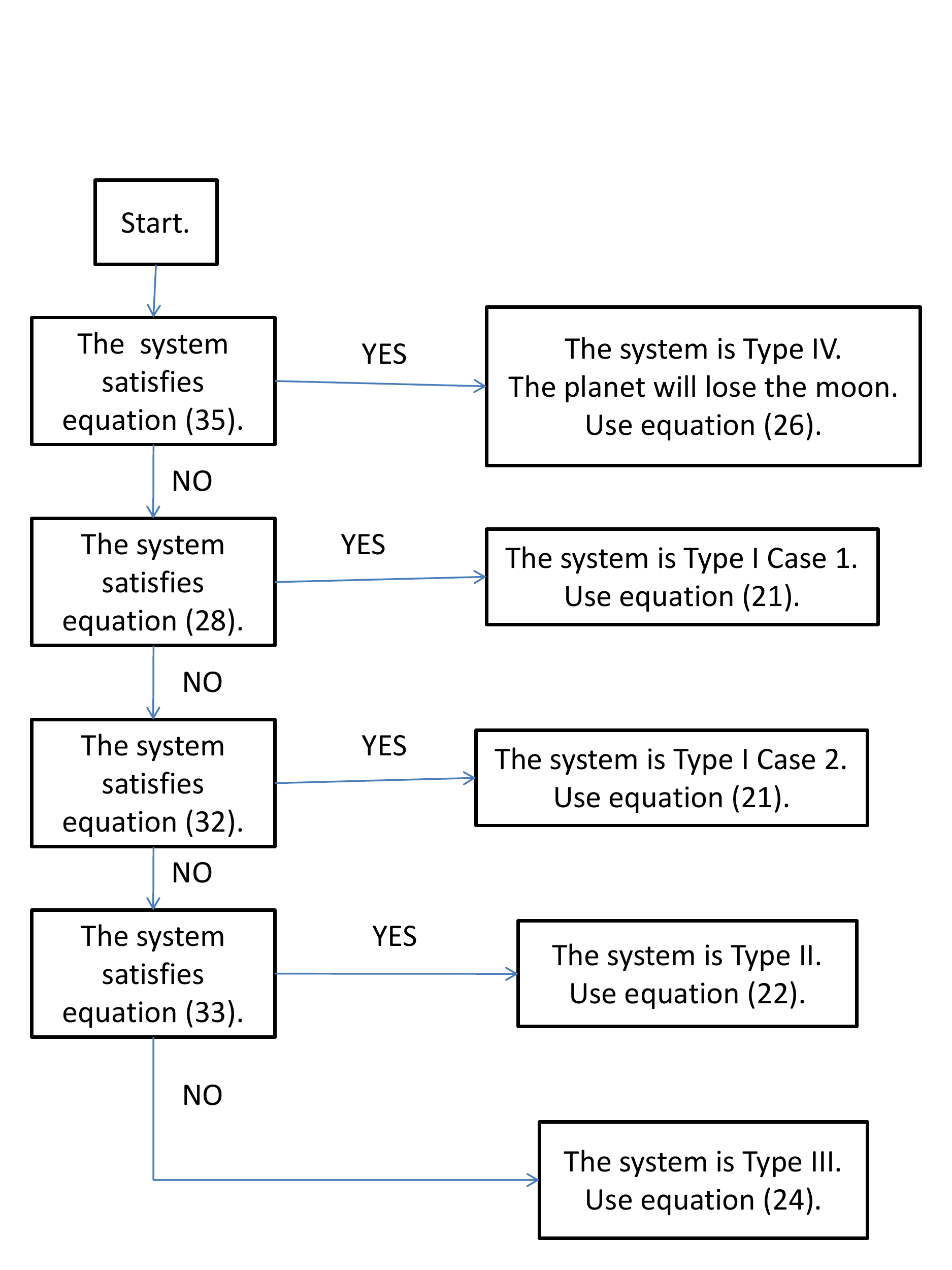}
\end{center}
\caption{Flow-chart for calculating moon lifetimes in a star-planet-moon system. First, check the type of system. Then, calculate the lifetime of the moon.}
\label{fig:sum}
\end{figure}

\section*{Appendix}
\subsection*{Type I Solution}
\begin{figure}[htb]
\begin{center}
\includegraphics[width=100mm]{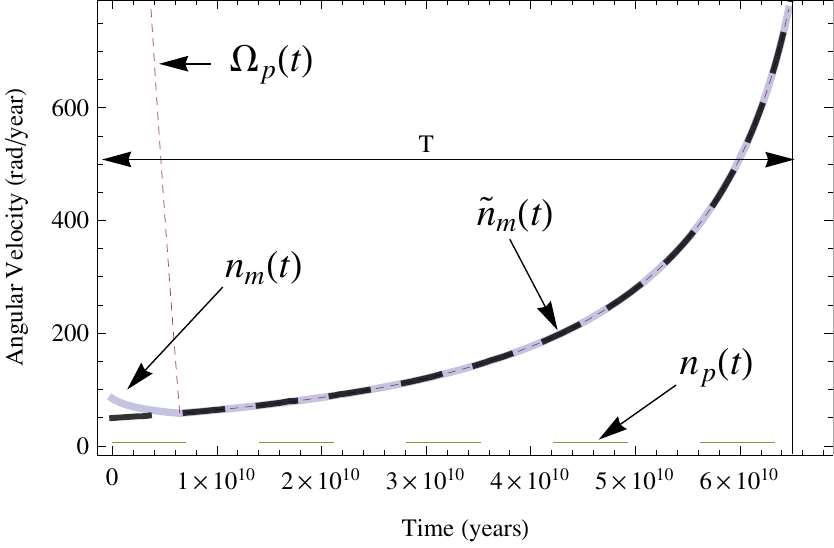}
\end{center}
\caption{The thick dashed line is $\widetilde{n}_{m}(t)$.
        The thick light blue line is $n_{m}(t)$.
        As you can see, $\widetilde{n}_{m}(t)$ and $n_{m}$ have the same maximum time.
        We use the present data of our Sun-Earth-Moon system.
        The initial conditions are $n_{m}(0)$=84 rad/year, $\Omega_{p}(0)=730 \pi$ rad/year, and $n_{p}(0)=2 \pi$ rad/year.
        The new initial conditions are $\widetilde{n}_{m}(0)$=$\Omega_{p}(0)=48.5524$ rad/year and $n_{p}(0)=2 \pi$ rad/year.}
\label{fig:newnm}
\end{figure}
Suppose that the initial conditions $n_{m}(0)$, $\Omega_{p}(0)$, and $n_{p}(0)$ are known. From these initial conditions, we can calculate $n_{m}(t)$, $\Omega_{p}(t)$, and $n_{p}(t)$ by solving the equations (\ref{eq:basiceq}) numerically (Fig.\ref{fig:TyIC1}).

For $0\leq t< T1$, we can use equation (\ref{eq:sol2}) with ${\rm sgn}(\Omega_{p}-n_{m})={\rm sgn}(\Omega_{p}-n_{p})=1$ because the planet is not tidally locked with either the star nor the moon, and $\Omega_{p}>n_{m}>n_{p}$. For $T1\leq t < T$, we can use (\ref{eq:sol2`}) with ${\rm sgn}(\Omega_{p}-n_{p})=1$ because the planet is tidally locked with the moon, and $\Omega_{p}>n_{p}$.

Define a function $\widetilde{{n}}_{m}(t)$ such that $\widetilde{{n}}_{m}(t)$ satisfies the following equation for $0\leq t<T$;
\begin{equation}
\frac{M_{m}(GM_{p})^{2/3}}{\widetilde{{n}}^{1/3}_{m}(t)}
+\alpha R_{p}^{2}M_{p}\widetilde{{n}}_{m}(t)+\frac{M_{p}(GM_{s})^{2/3}}{n_{p}^{1/3}(t)}
=L_{0}\nonumber
\end{equation}
where $L_{0}=\frac{M_{m}(GM_{p})^{2/3}}{n_{m}^{1/3}(0)}
+\alpha R_{p}^{2}M_{p}\Omega_{p}(0)+\frac{M_{p}(GM_{s})^{2/3}}{n_{p}^{1/3}(0)}$ is the initial angular momentum of the system.
In other words, $\widetilde{{n}}_{m}(t)$ represents situation in which the planet and the moon are tidally locked from beginning to end.
Because $\widetilde{{n}}_{m}(t)$ and $n_{m}(t)$ are the same for $T1 \leq t< T$, we can calculate the maximum lifetime of the moon if we know the domain of $\widetilde{{n}}_{m}(t)$ (Fig.\ref{fig:newnm}).

For given t, set $x=\widetilde{n}_{m}$ and define
\begin{equation}\label{eq100}
f(x)\equiv\frac{(GM_{m})(GM_{p})^{2/3}}{x^{1/3}}
+\alpha R_{p}^{2}(GM_{p})x+\frac{(GM_{p})(GM_{s})^{2/3}}{n_{p}(t)^{1/3}}-GL_{0}\nonumber.
\end{equation}
The condition that $f(x)$ has at least one zero, i.e. equation (\ref{eq:sol2`b}) has a real solution,
is
\begin{equation}
\frac{4}{3^{3/4}}\left\{(GM_{m})^3(GM_{p})^3\alpha R_{p}^2\right\}^{1/4}+\left(\frac{(GM_{p})(GM_{s})^{2/3}}{n_{p}(t)^{1/3}}-GL_{0}\right)\leq0.
\end{equation}
Since we know $n_{p}(t)$, we can plug in equation (\ref{eq:sol2`a}) and solve for $t$:
\begin{eqnarray}
{\scriptstyle
\left.
\begin{array}{ll}
t\leq\frac{2}{39}\frac{Q_{p}}{k_{2p}R_{p}^5}(GM_{p})(GM_{s})^{2/3}\\
\quad\left[\left(\frac{3^{3/4}GL_{0}-4\{(GM_{m})^3(GM_{p})^3\alpha R_{p}^2\}^{1/4}}{3^{3/4}(GM_{p})(GM_{s})^{2/3}}\right)^{13}-\left(\frac{1}{n_{p}(0)}\right)^{13/3}\right].
\end{array}
\right.
}
\end{eqnarray}
Hence, the lifetime of the moons for Type I, $T$, is
\begin{eqnarray}
{\scriptstyle
\left.
\begin{array}{ll}
T=\frac{2}{39}\frac{Q_{p}}{k_{2p}R_{p}^5}(GM_{p})(GM_{s})^{2/3}\\
\qquad\left[\left(\frac{3^{3/4}GL_{0}-4\{(GM_{m})^3(GM_{p})^3\alpha R_{p}^2\}^{1/4}}{3^{3/4}(GM_{p})(GM_{s})^{2/3}}\right)^{13}-\left(\frac{1}{n_{p}(0)}\right)^{13/3}\right].
\end{array}
\right.
}
\end{eqnarray}

\subsection*{Type II Solution}
For Type II, there are three stages (Fig.\ref{fig:T123}).
We start by finding $T2$.
Assume $n_{m}(T1)$ and $n_{p}(T1)$ are known.
At the end of the planet-star synchronized state, the torque due to the star is equal to the torque due to the planet.
Hence, $\tau_{p-s}(T1+T2)=\tau_{p-m}(T1+T2)$.
From equation (\ref{eq:eq1}), equation (\ref{eq:eq2}) and Kepler's Law,
\begin{equation}\label{eq:torqueT12}
n_{m}(T1+T2)=\left(\frac{GM_{p}}{GM_{m}}\right)^{1/2}n_{pc}
\end{equation}
where $n_{pc}=n_{p}(T1+T2)$ (Fig. \ref{fig:T123}).

For Stage 2, we can see that $\Omega_{p}-n_{m}<0$ from Fig. \ref{fig:T123}.
Hence, by the equation (\ref{eq:sol2``a}) with $n_{m}(0)=n_{m}(T1)$
\begin{displaymath}
n_{m}(t)=\left(-\frac{39}{2}\frac{k_{2p}R_{p}^{5}}{Q_{p}}\frac{Gm_{m}}{(GM_{p})^{8/3}}
             \:t+n_{m}^{-13/3}(T1)\right)^{-3/13}.
\end{displaymath}
But in this equation, we measure the time $t$ from $T1$.
Measuring the time $t$ from zero, for $T1\leq t \leq T2$,
\begin{displaymath}
n_{m}(t)=\left(-\frac{39}{2}\frac{k_{2p}R_{p}^{5}}{Q_{p}}\frac{Gm_{m}}{(GM_{p})^{8/3}}
             \:(t-T1)+n_{m}^{-13/3}(T1)\right)^{-3/13}.
\end{displaymath}

Hence,
\begin{equation}\label{eq:nmT12}
n_{m}(T1+T2)=\left(-\frac{39}{2}\frac{k_{2p}R_{p}^{5}}{Q_{p}}\frac{Gm_{m}}{(GM_{p})^{8/3}}
             \:T2+n_{m}^{-13/3}(T1)\right)^{-3/13}.
\end{equation}

For stage 1, we can see $\Omega_{p}-n_{m}>0$ from Fig. \ref{fig:T123}.
Hence, by the equation (\ref{eq:sol2a}),
\begin{equation}\label{eq:nmT1}
n_{m}(T1)=\left(\frac{39}{2}\frac{k_{2p}R_{p}^{5}}{Q_{p}}\frac{Gm_{m}}{(GM_{p})^{8/3}}
             \:T1+n_{m}^{-13/3}(0)\right)^{-3/13}.
\end{equation}

Combine equation (\ref{eq:torqueT12}), equation (\ref{eq:nmT12}), and equation (\ref{eq:nmT1}).  Solve for $T2$.
\begin{eqnarray}\label{eq:T2}
{\scriptstyle
T2=T1+\frac{2}{39}\frac{Q_{p}}{k_{2p}R_{p}^5}\frac{(GM_{p})^{8/3}}{(GM_{m})}
    \left[n_{m}(0)^{-13/3}-\left\{\left(\frac{GM_{p}}{GM_{m}}\right)^{1/2}n_{pc}\right\}^{-13/3}\right]}.
\end{eqnarray}

Next, we will find $n_{pc}$ and $T3$.
At $t=T1+T2$, we can see that $\Omega(t)=n_{p}(t)=n_{pc}$ from Fig.\ref{fig:T123}.  By the conservation of angular momentum and equation (\ref{eq:torqueT12}),
\begin{equation}
GL_{0}=\frac{(GM_{m})^{7/6}(GM_{p})^{1/2}+(GM_{p})(GM_{s})^{2/3}}{n_{pc}^{1/3}}
+\alpha R_{p}^{2}(GM_{p})n_{pc}.
\end{equation}
The second term, $\alpha R_{p}^{2}(GM_{p})n_{pc}$, is the spin angular momentum of the planet.
We will ignore this term to approximate $n_{pc}$ because we know the spin angular momentum is small compared to the total angular momentum.

Hence,
\begin{equation}\label{eq:npc}
n_{pc}\approx\left\{\frac{(GM_{m})^{7/6}(GM_{p})^{1/2}+(GM_{p})(GM_{s})^{2/3}}{GL_0}\right\}^3.
\end{equation}

When we use $\Omega_{p}(0)=n_{p}(0)=n_{pc}$ and $n_{m}(0)=n_{m}(T1+T2)$ as a initial condition, we can calculate $T3$ by using the formula for Type I.  Hence,
\begin{eqnarray}\label{eq:T3}
{\scriptstyle
\begin{array}{ll}
T3=\frac{2}{39}\frac{Q_{p}}{k_{2p}R_{p}^5}(GM_{p})(GM_{s})^{2/3}\\
\qquad\left[\left(\frac{3^{3/4}GL_{0}-4\{(GM_{m})^3(GM_{p})^3\alpha R_{p}^2\}^{1/4}}{3^{3/4}(GM_{p})(GM_{s})^{2/3}}\right)^{13}-\left(\frac{1}{n_{pc}}\right)^{13/3}\right].
\end{array}
}
\end{eqnarray}

From equation (\ref{eq:T2}), equation (\ref{eq:npc}), and the equation (\ref{eq:T3}), the total life time, $T$, is
\begin{eqnarray}
{\scriptstyle
\begin{array}{ll}
T=T1+T2+T3\\
 \quad=2T1+\frac{2}{39}\frac{Q_{p}}{k_{2p}R^5_{p}}
 \left[
 \frac{(GM_{p})^{8/3}}{(GM_{m})}n^{-13/3}_{m}(0)\right.\\
\qquad +\left.\frac{\left(3^{3/4}GL_{0}-4\left\{(GM_{m})^3(GM_{p})^3\alpha R_{p}^2\right\}^{1/4}\right)^{13}}{3^{39/4}(GM_p)^{12}(GM_{s})^{8}}\right.\\
 \qquad\left.-\frac{(GL_0)^{13}}{\left\{(GM_{p})^{1/2}(GM_{m})^{7/6}+(GM_{p})(GM_{s})^{2/3}\right\}^{12}}\right].
\end{array}
}
\end{eqnarray}

\subsection*{Type III Solution}
In Type III, the graph of $n_{m}(t)$ is comprised of two parts. The first part, $n^+_{m}(t)$, is from $0\leq t\leq T1$ and the second part, $n^-_{m}(t)$ is from $T1\leq t<T$.
sgn($\Omega_{p}-n_{m}$) is 1 for $0\leq t\leq T1$ and -1 for $T1\leq t<T$ because the planet and the moon are not in a synchronized state. To have $n^+_{m}(t)$, we can use equation (\ref{eq:sol2}) directly. To have $n^-_{m}(t)$, we set $n_{m}(0)=n_{m}(T1)$ and $t=t-T1$ because we need to shift the graph $T1$ in a positive direction.
From equation (\ref{eq:sol2}),
\begin{eqnarray}\label{eq:nm+}
n^+_{m}(t)=\left(\frac{39}{2}\frac{k_{2p}R_{p}^{5}}{Q_{p}}\frac{GM_{m}}{(GM_{p})^{8/3}}\:t \:+n_{m}^{-13/3}(0)\right)^{-3/13}
\end{eqnarray}
for $0<t<T1$ and
\begin{eqnarray}\label{eq:nm-}
&&n^-_{m}(t)=\left(-\frac{39}{2}\frac{k_{2p}R_{p}^{5}}{Q_{p}}\frac{GM_{m}}{(GM_{p})^{8/3}}\:(t-T1)\:\right.\\\nonumber
&&\qquad\qquad\qquad\qquad\qquad\qquad\qquad\left.+\left\{n^+_{m}(T1)\right\}^{-13/3}\right)^{-3/13}
\end{eqnarray}
for $T1<t<T$.

By the symmetry, $Tb=T1+\frac{n_{m}^{-13/3}(0)}{a}$ where $a=\frac{39}{2}\frac{k_{2p}R_{p}^{5}}{Q_{p}}\frac{GM_{m}}{(GM_{p})^{8/3}}$.  Hence,
\begin{eqnarray}
T&=&T1+Tb\nonumber\\
 &=&2T1+\frac{n_{m}^{-13/3}(0)}{a}\nonumber\\
 &=&2T1+\frac{2}{39}\frac{Q_{p}}{k_{2p}R_{p}^{5}}\frac{(GM_{p})^{8/3}}{GM_{m}}{n_{m}^{-13/3}(0)}.
\end{eqnarray}
We can get the same result if we set the inside of the parenthesis of equation (\ref{eq:nm-}) equal to 0, and then solve for $t$.

\subsection*{Condition for Type I Case 1}
The condition for Type I Case 1 is that the magnitude of the torque due to the moon is greater than the magnitude of the torque due to the star at $t=T1$ (Fig.\ref{fig:TyIC1}),
\begin{equation}
|\tau_{p-m}(T1)|\geq|\tau_{p-s}(T1)|.
\label{TIC1}
\end{equation}
By equation (\ref{eq:eq1}), equation (\ref{eq:eq2}), and Kepler's Law, equation (\ref{TIC1}) implies that
\begin{equation}\label{eq:torques}
n_{m}(T1)\geq\left(\frac{GM_p}{GM_m}\right)^{\frac{1}{2}}n_{p}(T1).
\label{nmnp}
\end{equation}
For the time period $0\leq t \leq T1$, the planet is not tidally locked with either the star or the moon.
We can use equations (\ref{eq:sol2}) with sgn$(\Omega_{p}-n_{m})$=1 and sgn$(\Omega_{p}-n_{p})$=1.

Combine equation (\ref{nmnp}), equation (\ref{eq:sol2a}), and equation (\ref{eq:sol2b}), then solve for $T1$. We obtain
\begin{eqnarray}
T1\leq\frac{2}{39}\frac{Q_{p}}{k_{2p}R_{p}^{5}}\frac{(GM_{p})(GM_{m})^{7/6}(GM_{s})^{2/3}}{(GM_{p})^{1/2}(GM_{s})^{2/3}-(GM_{m})^{7/6}}\nonumber\\
\times\left\{n_{p}^{-13/3}(0)-\left(\frac{GM_{p}}{GM_{m}}\right)^{13/6}n_{m}^{-13/3}(0)\right\}.
\end{eqnarray}

\subsection*{Condition for Type I Case 2}
Assume $n_{p}(T1)$ and $n_{m}(T1)$ are known.  Let $t_{*}$ be the time from $T1$, when the magnitudes of two torques are equal (Fig.\ref{fig:TIC2mag}).

The condition for Type I Case 2 is
\begin{equation}\label{TIC2}
\Omega_{p}(t_{*})\geq n_{p}(t_{*})
\end{equation}
where $t_{*}$ satisfies
\begin{equation}\label{eq:taupmps}
|\tau_{p-m}(t_{*})|=|\tau_{p-s}(t_{*})|.
\end{equation}
Equation (\ref{eq:taupmps}) implies that
\begin{equation}
n_{m}(t_{*})=\left(\frac{GM_p}{GM_m}\right)^{1/2}n_{p}(t_{*}).
\label{nm*}
\end{equation}
We did a similar calculation to arrive at equation (\ref{eq:torques}).

By conservation of angular momentum and equation (\ref{nm*}), we derive
\begin{eqnarray}
\scriptscriptstyle{
\begin{array}{ll}
\Omega_{p}(t_{*})=\frac{1}{\alpha R^2_{p} (GM_{p})}\\
\qquad\left[GL_{0}-\frac{(GM_{p})^{1/2}\{(GM_{m})^{7/6}+(GM_{p})^{1/2}(GM_{s})^{2/3}\}}{n^{1/3}_{p}(t_{*})}\right].
\end{array}
}\label{Omega}
\end{eqnarray}

After $T1$, the planet is not tidally locked with either the star or the moon for a time.
We can use equations (\ref{eq:sol2}) with sgn$(\Omega_{p}-n_{m})=-1$ and sgn$(\Omega_{p}-n_{p})=1$.
We use equation (\ref{eq:sol2a}) and equation (\ref{eq:sol2b}) with initial conditions $n_{m}(0)=n_{m}(T1)$ and $n_{p}(0)=n_{p}(T1)$ , which are
\begin{equation}\label{eq:nmt1t1t2}
n_{m}(t)=\left(-\frac{39}{2}\frac{k_{2p}R_{p}^{5}}{Q_{p}}\frac{GM_{m}}{(GM_{p})^{8/3}}
             \:t+n_{m}^{-13/3}(T1)\right)^{-3/13}
\end{equation}
and
\begin{equation}\label{eq:npt1t1t2}
n_{p}(t)=\left(\frac{39}{2}\frac{k_{2p}R_{p}^{5}}{Q_{p}}\frac{1}{(GM_{p})(GM_{s})^{2/3}}
             \:t+n_{p}^{-13/3}(T1)\right)^{-3/13}.
\end{equation}

Plug in equation (\ref{eq:nmt1t1t2}) and equation (\ref{eq:npt1t1t2}) into the equation (\ref{nm*}), and solve for $t_{*}$. Then, we have
\begin{eqnarray}
\scriptscriptstyle{
\begin{array}{ll}
t_{*}=\frac{39}{2}\frac{Q_{p}}{k_{2p}R_{p}^{5}}\frac{(GM_{p})(GM_{m})^{7/6}(GM_{s})^{2/3}}{\{(GM_{m})^{7/6}+(GM_{p})^{1/2}(GM_{s})^{2/3}\}}\\
\qquad\qquad\times\left\{\left(\frac{GM_{p}}{GM_{m}}\right)^{13/6}n_{m}^{-13/3}(T1)-n_{p}^{-13/3}(T1)\right\}.
\end{array}
}
\end{eqnarray}

By using this $t_{*}$, we have
\begin{eqnarray}
&&\Omega_{p}(t_{*})=c(GL_0-a_1 b^{12} X)\\
&&n_{p}(t_{*})=a_2 b^3 X^{-3}
\end{eqnarray}
where
\begin{eqnarray}
\small{
\begin{array}{ll}
c=\frac{1}{\alpha R^2_{p} (GM_p)}\nonumber\\
a_1=(GM_p)^{1/2}(GM_m)^{7/78}\nonumber\\
a_2=\frac{1}{(GM_m)^{7/26}}\nonumber\\
b=\left\{(GM_{m})^{7/6}+(GM_{p})^{1/2}(GM_{s})^{2/3}\right\}^{1/13}\nonumber\\
X=\left\{\left(\frac{GM_{p}}{GM_{m}}\right)^{13/6}n_{m}^{-13/3}(T1)+\frac{(GM_p)^{1/2}(GM_s)^{2/3}}{(GM_m)^{7/6}}n_{p}^{-13/3}(T1)\right\}^{1/13}.
\end{array}
}
\end{eqnarray}

Applying equation (\ref{TIC2}), we have
\begin{eqnarray}
a_1 b^{12} X^4-GL_0 X^3+\frac{a_2}{c}b^3\leq0.
\end{eqnarray}

\subsection*{Conditions for Type II and III}
From Fig.\ref{fig:ConIII} and Fig.\ref{fig:ConII}, the condition for Type II is $T5>T6$ and the condition for Type III is $T5\leq T6$.

For Stage 3, we know
\begin{eqnarray}\label{eq:solstIII}
\left\{
\begin{array}{ll}
n_{m}^{'}(t)=\left(-\frac{39}{2}\frac{k_{2p}R_{p}^{5}}{Q_{p}}\frac{GM_{m}}{(GM_{p})^{8/3}}\:t\right.\\
             \qquad\qquad\qquad\qquad\qquad\left.+n_{m}^{-13/3}(T1+T2)\right)^{-3/13}\\
n_{p}(t)=\left(\frac{39}{2}\frac{k_{2p}R_{p}^{5}}{Q_{p}}\frac{1}{(GM_{p})(GM_{s})^{2/3}}\:t\right.\\
             \qquad\qquad\qquad\qquad\qquad\left.+n_{p}^{-13/3}(T1+T2)\right)^{-3/13}.\\
\end{array}
\right.
\end{eqnarray}

At $t=T6$, $n_{m}^{'}(t)=\infty$.  Hence,
\begin{equation}
T6=\frac{2}{39}\frac{Q_{p}}{k_{2p}R_{p}^{5}}\frac{(GM_{p})^{8/3}}{GM_{m}}n_{m}^{-13/3}(T1+T2).
\end{equation}

To find $T5$, we use the same formula for Type I, which is equation (\ref{eqLFTI}).
We set $n_{p}(0)=n_{p}(T1+T2)$.
$T5$ is
\begin{eqnarray}
\left.
\begin{array}{ll}
T5=\frac{2}{39}\frac{Q_{p}}{k_{2p}R_{p}^5}(GM_{p})(GM_{s})^{2/3}\times\\
\left[\left(\frac{3^{3/4}GL_{0}-4\{(GM_{m})^3(GM_{p})^3\alpha R_{p}^2\}^{1/4}}{3^{3/4}(GM_{p})(GM_{s})^{2/3}}\right)^{13}-\left(\frac{1}{n_{p}(T1+T2)}\right)^{13/3}\right].
\end{array}
\right.
\end{eqnarray}

From (Fig.\ref{fig:T123}), $n_{pc}=n_{p}(T1+T2)$. And we know that $n_{m}(T1+T2)=(\frac{GM_{p}}{GM_{m}})^{1/2}n_{pc}$ from equation (\ref{eq:torqueT12}).  By using this information, we have
\begin{eqnarray}
\left\{
\begin{array}{ll}
T5=\frac{2}{39}\frac{Q_{p}}{k_{2p}R_{p}^5}(GM_{p})(GM_{s})^{2/3}\\
\qquad\times\left[\left(\frac{3^{3/4}GL_{0}-4\{(GM_{m})^3(GM_{p})^3\alpha R_{p}^2\}^{1/4}}{3^{3/4}(GM_{p})(GM_{s})^{2/3}}\right)^{13}-n_{pc}^{-13/3}\right]\\
T6=\frac{2}{39}\frac{Q_{p}}{k_{2p}R_{p}^{5}}\frac{(GM_{p})^{1/2}}{GM_{m}^{7/6}}n_{pc}^{-13/3}.
\end{array}
\right.
\end{eqnarray}
From equation (\ref{eq:npc}), we also know that
\begin{equation}
n_{pc}=\left\{\frac{(GM_{m})^{7/6}(GM_{p})^{1/2}+(GM_{p})(GM_{s})^{2/3}}{GL_0}\right\}^3.
\end{equation}

The condition for Type II, $T5>T6$, implies
\begin{eqnarray}\label{eq:TII}
{\scriptstyle
\begin{array}{ll}
 \left(3^{3/4}GL_{0}-4\left\{(GM_{m})^3(GM_{p})^3\alpha R_{p}^2\right\}^{1/4}\right)^{13}\times\\
\left((GM_{m})^{7/6}+(GM_{p})^{1/2}(GM_{s})^{2/3}\right)^{12}\\
> 3^{39/4}(GM_{p})^6(GM_{s})^8(GL_{0})^{13}.
\end{array}
}
\end{eqnarray}

Similarly, the condition for Type III, $T5\leq T6$, implies
\begin{eqnarray}\label{eq:TIII}
{\scriptstyle
\begin{array}{ll}
 \left(3^{3/4}GL_{0}-4\left\{(GM_{m})^3(GM_{p})^3\alpha R_{p}^2\right\}^{1/4}\right)^{13}\times\\
\left((GM_{m})^{7/6}+(GM_{p})^{1/2}(GM_{s})^{2/3}\right)^{12}\\
\leq 3^{39/4}(GM_{p})^6(GM_{s})^8(GL_{0})^{13}.
\end{array}
}
\end{eqnarray}

 \section*{Acknowledgements}
 TS and JWB are supported by the NASA Exobiology program grant NNX09AM99G and DPO is supported by a NASA Fellowship for Early Career Researchers with the grant number NNX10AH49G.
\bibliographystyle{apj}
\bibliography{myreferences}

\end{document}